\documentclass[prb,aps,epsfig,showpacs,superscriptaddress,twocoloumn,15point,reprint]{revtex4-2}
\usepackage{graphicx}
\usepackage{dcolumn}
\usepackage{bm}
\makeatletter
\usepackage{amsmath}
\usepackage{epstopdf}
\usepackage{lineno,hyperref}
\usepackage{array}

\newcommand{\Rmnum}[1]{\expandafter\@slowromancap\romannumeral #1@}
\makeatother
\begin{document}
\title{\bf Critical Analysis of Skyrmionic Material Co$_{6.5}$Ru$_{1.5}$Zn$_8$Mn$_4$: a complex interplay of short and long-range interactions around the transition temperature}

\author{Afsar Ahmed  }
\author{Arnab Bhattacharya }
\author{ Samik DuttaGupta}
\author{ I. Das}
\affiliation{CMP Division, Saha Institute of Nuclear Physics, A CI of Homi Bhabha National Institute, Kolkata 700064, India}

\begin{abstract}
Critical behaviour study in magnetism is important owing to its application for understanding the nature of underlying spin-spin interactions by determining the critical parameters in the vicinity of a phase transition. In this article, we report the novel manifestation of crossover behaviour between two universality classes governing spin interaction across the ferromagnetic Curie temperature $T_C$ in critical scaling of anomalous hall conductivity isotherms for a skyrmion-hosting itinerant ferromagnet Co$_{6.5}$Ru$_{1.5}$Zn$_{8}$Mn$_4$. Along with the magnetotransport scaling, the traditional critical behaviour of magnetic isotherms yields $\beta$ = 0.423 $\pm$ 0.004, $\gamma$ = 1.08 $\pm$ 0.016, and $\delta$ = 3.553 $\pm$ 0.009 suggesting the 3D Heisenberg and Mean field type of spin interactions below and above $T_C$, respectively. The isotropic magnetic exchange strength decays as $J(r) \approx r^{ -4.617}$, implying the prevalence of crossover from long-range ordering to short-range type interaction. In addition, the existence of a fluctuation-disordered magnetic phase immediately below $T_C$ has been observed in the magnetocaloric effect. The novel approach of generating a low-field phase diagram employing the quantitative criterion of phase transition from the scaling of isothermal magneto-entropic change shows an excellent convergence with the phase boundaries obtained from conventional magnetic and anomalous Hall conductivity scaling. This simultaneous scaling of magnetization and AHC isotherms for systems with crossover behaviour establishes the universality of the magnetotransport-based critical scaling approach which still remains in its infancy.
\end{abstract}

\maketitle

\section{Introduction}

Critical behaviour study in simple ferromagnets has always been of great interest for condensed matter research to understand the nature of underlying interaction mechanisms leading to the establishment of magnetic order. The recent emergence of various complex topologically protected magnetic phases as a phase pocket just below the ferromagnetic transition temperature ($T_C$), has brought back the critical behaviour study to the limelight for understanding the spin exchange interaction mechanism leading to the manifestation of unconventional spin textures \cite{MnSic, FeGec, FeCoSic,Cu2SeO3c, GaV4S8c}. Among the various family of spin structures in ferromagnetic systems, skyrmion (Skx) have attracted a lot of attention due to its unconventional electrodynamics and potential for application in high-density information storage\cite{app1MnSi, app2, app3, app4, app5, app6, app7cu2seO3}, and has triggered extensive research for high $T_C$ materials with skyrmionic $A$-phase near the transition temperature, in bulk as well as hetero-structure thin films \cite{centsymm, gd2pdSi3, app1MnSi, FeGe, FeCoSi, cu2seO3, app7cu2seO3, FeGefilm}. For each of these materials, the underlying spin interaction mechanism stabilizing the Skx phase ranges from long-range mean-field or distance-dependant RKKY \cite{gd2pdSi3, gdru2si2,lim2020emergent, chowdhury2021unconventional, chowdhury2022modification} to short-range 3D Heisenberg interaction. To create an $A$-phase in the vicinity of $T_C$, the competition between symmetric Heisenberg [\textit{-J($\vec{S_i}.\vec{S_j}$)}] and asymmetric Dzyaloshinskii-Moriya interaction [\textit{-D($\vec{S_i}\times\vec{S_j}$)}], where $\vec{S_i}$ and $\vec{S_j}$ are interacting spins, plays a crucial role. The symmetric interaction $J$ provides an opportunity to tune the size and stability of the skyrmionic lattice, entailing a deeper understanding of the underlying spin interaction for the design of new exotic magnetic-phase hosting material systems.

Recently, Tokura et. al\cite{coznmn} have discovered a new class of structurally disordered intermetallic compounds Co$_x$Zn$_x$Mn$_x$ (x+y+z = 20) with high transition temperature, hosting complex magnetic noncoplanar structures like square lattice skyrmion\cite{karube2016robust}, hexagonal skyrmionic lattice \cite{tokunaga2015new}, chiral solitonic lattice \cite{karubecoznmn,yu2018transformation}, and elongated skyrmions\cite{morikawa2017deformation} with conical and helical phases just below $T_C$ and a reentrant spin glassy behaviour at low temperature. The intermetallic compound Co$_8$Zn$_8$Mn$_4$\cite{menzel2019local}, belonging to this class, and doping of the parent compound with Fe at the $8c$ lattice sites resulting in Co$_{3.5}$Fe$_{4.5}$Zn$_8$Mn$_8$\cite{bhattacharya2023critical}, have been reported to show a short-range interaction which can be classified into 3D Heisenberg universality class. This doping also in turn dilutes the DM interaction strength and further shifts the position of the SkL (skyrmion lattice) pocket to lower values along the $H$ axis in the phase diagram. Unlike the above behaviour, the strength of DMI remains fairly constant\cite{Fecoznmn} on replacing the Co atoms with $4d$ Ruthenium atoms in comparison to Fe atoms. This occurs due to a smaller exchange splitting for $4d$ bands which subsequently change the transition temperature $T_C$ suggesting a modification in the underlying spin interaction mechanism for the Ru-substitution of Co in Co$_8$Zn$_8$Mn$_4$. This makes Co$_{6.5}$Ru$_{1.5}$Zn$_8$Mn$_4$ system as an interesting candidate to study the effect of intermixing the $3d$ and $4d$ bands and its effect on spin interaction near $T_C$.

Here, we study the critical behaviour of Co$_{6.5}$Ru$_{1.5}$Zn$_{8}$Mn$_4$ using the conventional magnetic isotherm approach and the magnetotransport-based scaling. We observe a crossover behaviour between two universality classes of spin interactions \textit{i.e.} from short-range Heisenberg to long-range Mean field, below and above $T_C$, respectively. Thus making our system unique from the earlier reports of spin interaction mechanisms for compounds belonging to the same family\cite{bhattacharya2023critical,menzel2019local}. However, in a few rare materials, the spin interactions below and above $T_C$ cannot be attributed to a single universality class but show a crossover behaviour between two different mechanisms \cite{Cu2SeO3c,pramanik2009critical,taran2005critical,poon1977critical,MnV2O4} which arises due to many reasons i.e dilution of global spin by doping of non-magnetic ions, first-order structural transition, phase separation, fluctuation of the order parameter etc.. The thermodynamics scaling approach was also employed to understand the dependency of $\Delta S_m$ on external applied magnetic field $H$ (i.e $\Delta S_m$ $\propto$ $H^n$) and the temperature and field dependency of n was used to chalk out the phase boundaries in the critical temperature region. We observe an excellent agreement of the obtained universality classes representing the spin interactions across $T_C$, using all of the above mentioned approaches. Using the obtained results, we constructed the magnetic phase diagram by using the quantitative criterion for phase transition\cite{law2018quantitative}. The obtained results show the possibility of using magnetic and magnetotransport-based scaling approaches as an effective tool to understand the nature of magnetic interactions in systems with complex magnetic ordering.

\section{Experimental details}

Polycrystalline Co$_{6.5}$Ru$_{1.5}$Zn$_{8}$Mn$_4$ was prepared using solid-state reaction technique, similar to previous reports \cite{Fecoznmn, xie2013beta, characterization}. The phase purity of the sample was checked by X-ray diffraction (XRD) at room temperature, and the observed XRD patterns were indexed using Rietveld refinement. The elemental composition and homogeneity of the samples were confirmed by a transmission electron microscope (TEM) equipped with an energy-dispersive X-ray (EDX) spectrometer. Magnetic properties were characterized using a superconducting quantum interference device (SQUID-VSM), and physical property measurement system (PPMS) in the temperature (\textit{T}) range of 2-300 K, using samples of approximate dimensions 2.67 mm $\times$ 1.15 mm $\times$ 0.45 mm. The magnetotransport measurements were performed by a 9 T physical property measurement system (PPMS), in the \textit{T} range of 2-300 K using a standard four-probe arrangement, under applied dc of 10 mA. Before measurements, the ohmic nature of the contacts was checked over the entire \textit{T} range. Magnetization (\textit{M}) versus \textit{T} measurements were carried out in two different protocols, ZFCW (zero field cooled warming) and FCC (field cooled cooling). Magnetic and magnetotransport isotherms were measured after cooling down the system from well above the transition temperature ($T_C$) to the desired \textit{T} under ZFCW protocol and rested for three minutes before each measurement. The transverse resistivity ($\rho_{xy}$) was symmetrized using the relation, $\rho_{xy} = \left[\rho_{xy}(+H) - \rho_{xy}(-H)\right]/2$. The effective magnetic field on the sample was calculated by $H = H_{app}-N_d M$, where $N_d$ = 1 is the demagnetization factor considering the plate-like geometry of the samples. These corrected values of the magnetic field (\textit{H}) were subsequently used for all calculations.

\begin{figure*}[t]
\begin{center}
\includegraphics[width=1\textwidth]{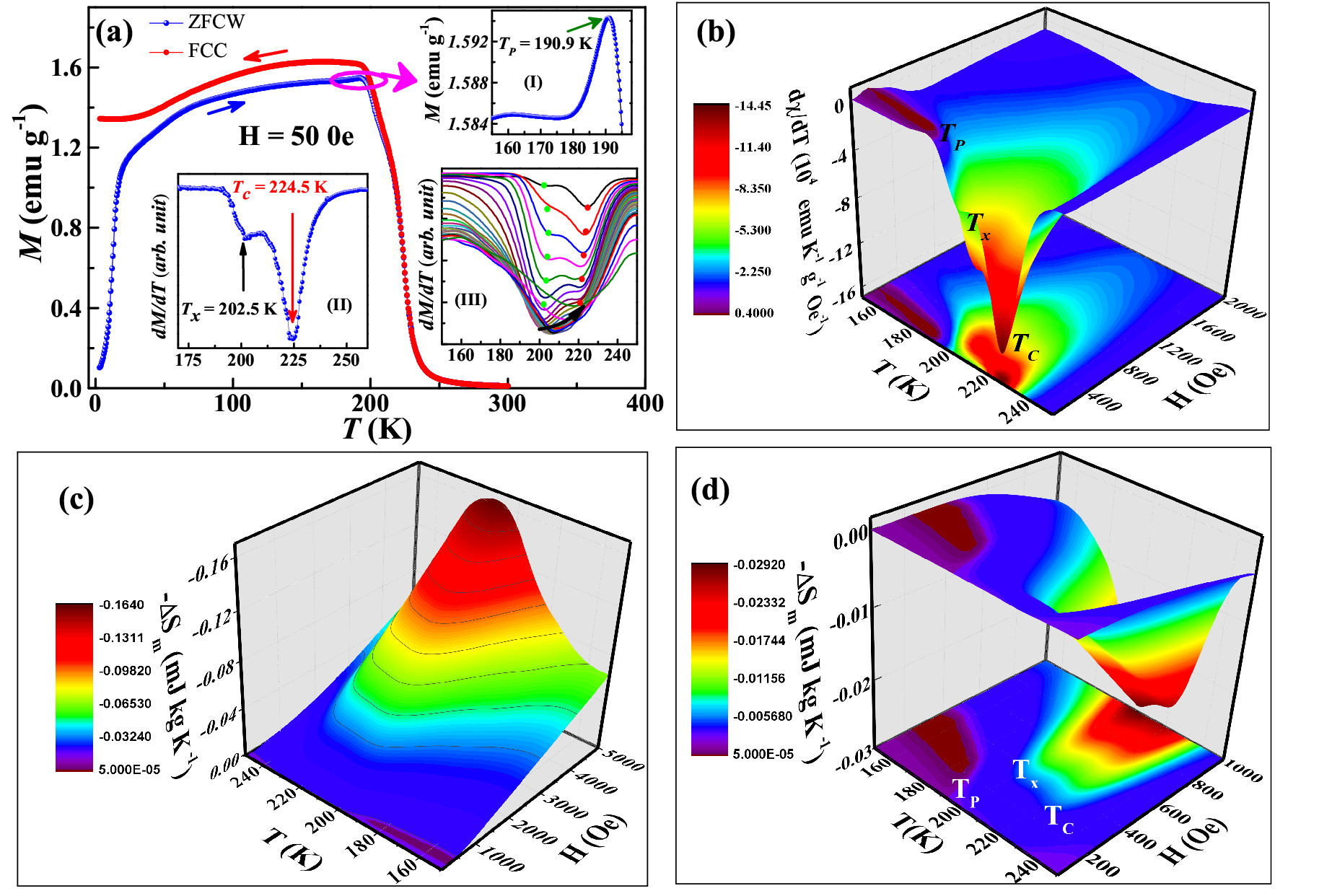}
\end{center}
\caption[]
{\label{MCE}(a) Magnetization (\textit{M}) vs. temperature (\textit{T}) at 50 Oe applied field in ZFCW (blue) and FCC (red) protocols. Inset (I) shows a cusp at around 190.9 K indicating an onset of helical ordering. The inset (II) shows $dM/dT$ vs. $T$ of 50 Oe ZFCW ( two minima observed at 202.5 K and 224.5 K represent two different magnetic transitions). Inset (III) shows the variation of the first-order derivative of Magnetization (dM/dT) versus temperature (\textit{T}) at the different applied field: red dotted line indicates slight shifting of $T_C$ at lower temperature and green dotted line represents shifting of $T_X$ at the higher temperature slightly as the applied field increases (below 800 Oe); at the sufficiently higher applied magnetic field, the overall minima shifts towards the higher temperature (black arrow) (b) The temperature and field dependence of d$\chi/dT$. The positive value of d$\chi/dT$ up to 400 Oe below $T_P$ and lowest value of $d\chi/dT$ is observed at the $T_C$ (black spot) along with a neck in between $T_P$ and $T_C$. (c) The temperature and field dependence of $\Delta$$S_m$. Except below $T_P$ and low \textit{H} ($<$ 400 Oe), $\Delta S_m$ is negative with a minimum near $T_C$, (d) Zoom view of the temperature and field dependence of $\Delta$$S_m$ (below 1000 Oe).}
\end{figure*}

\begin{table}[t]
\begin{center}
\caption{Crystallographic parameters of full Reitveld analysis of the compound $Co_{6.5}Ru_{1.5}Zn_{8}Mn_{4}$ where the number in parenthesis are the uncertainty in the refinement.  \label{XRD}}
\begin{tabular}{p{3.0 cm}p{2.5 cm}p{2.5 cm}}
\hline
\hline
\noalign{\smallskip}
Space Group & P2$_1$3 (Group No. 213)\\
Lattice parameters: & $a =$ 6.34(9)$\AA$  \\
$R_{f} = $& $11.95$ \\
$R_{Bragg} =$ & $8.35$ \\
$\chi^2 =$  & 1.2 \\
\noalign{\smallskip}
\noalign{\smallskip}
&\textbf{Atomic}&\\
& \textbf{Coordinates}&\\
\end{tabular}
\begin{tabular}{p{1.3 cm}p{1.675 cm}p{1.675 cm}p{1.675 cm}p{1.675 cm}}
Atom  &Wyckoff Position &  x   &   y  &   z  \\
\hline
\noalign{\smallskip}
Co & 8c & 0.0693(1) & 0.0693(1) & 0.0693(1) \\
Ru & 8c & 0.0693(1) & 0.0693(1) & 0.0693(1) \\
Mn1 & 8c & 0.0693(1) & 0.0693(1) & 0.0693(1) \\
Mn2 & 12d & 0.1250(1) & 0.2025(5) & 0.4525(5) \\
Zn & 12d & 0.1250(1) & 0.2025(5) & 0.4525(5) \\
\noalign{\smallskip}
\noalign{\smallskip}
\hline
\hline
\end{tabular}
\end{center}
\end{table}

The room temperature XRD pattern was analyzed by Rietveld refinement using FullProf software (see supplementary information) where all the peaks were indexed considering the chiral cubic space group P4$_1$32 (No. 213). Table \ref{L} shows the results of the obtained fitting parameters indicating a very good agreement with the previous report \cite{Fecoznmn}. The absence of any unindexed or additional peaks in the XRD pattern confirms the phase purity of the sample. The chemical composition was studied using EDX spectroscopy, confirming the atomic ratio of Co: Ru: Zn: Mn to 6.47(2): 1.53(4): 8.05(7): 3.98(3), within experimental accuracy (see supplementary information). \text{Figure} \ref{MCE}(a) shows the \textit{T} dependence of magnetization ($M(T)$) under an applied \textit{H} = 50 Oe, for both ZFCW and FCC conditions. For both, a ferromagnetic to paramagnetic phase transition is observed, associated with a smooth decrease of $M$ (around $T_C$) with an increase in \textit{T}. The ferromagnetic transition temperature $T_C \approx$ 224.5 K is determined from the minima of $\frac{dM}{dT}$ versus $T$, shown in the inset (II) of \text{Fig.} \ref{MCE}(a). With lowering of \textit{T}, a second minima is observed at 202.5 K (marked as $T_X$ in the inset (II) of \text{Fig.} \ref{MCE}(a)), along with a magnetization maxima present at $T (= T_P)=$ 190.9 K as shown in the inset (I) of \text{Fig.} \ref{MCE}(a). The presence of a maximum in the vicinity of $T_C$ is a well-observed behaviour in the parent class of compound \textit{i.e.}, Co$_x$Zn$_y$Mn$_z$ (x+y+z = 20) \cite{karubecoznmn} and other systems such as MnSi \cite{MnSic}, Fe$_{0.8}$Co$_{0.2}$Si \cite{FeCoSic} and Cu$_2$OSeO$_3$ \cite{Cu2SeO3c}, attributed to the stabilization of a non-collinear magnetic phase (\textit{i.e.}, helical or conical) under small external \textit{H}, manifesting in an increase of the net in-plane \textit{M}. The variation of longitudinal resistivity ($\rho_{xx}$) versus \textit{T} without any applied magnetic field is shown in the supplementary. Our experimental result shows a very weak temperature dependence above $T_C$ followed by a metal-insulator transition, similar to Fe$_{0.8}$Co$_{0.2}$Si \cite{manyala2000addendum}. To understand the itinerant or the localized nature of the ferromagnetic ordering, we performed the Rhodes-Wohlfarth analysis\cite{rhodes1963effective}. According to this theory, if $q_C/q_S$ $\sim$ 1, the nature of magnetic order is localized and for $q_C/q_S$ $>$ 1 it follows the itinerant theory, where $q_C$ is the number of magnetic carriers per atom and $q_S$ is the saturation moment. The effective paramagnetic moment $p_{eff}$ is determined from the slope of $\chi_0^{-1}$ vs \textit{T} curve using the Curie-Weiss law $\chi = \frac{C}{T - \theta}$ where \textit{C} = $N_A\mu_B^2p_{eff}^2/(3k_B)$, where $N_A$, $\mu_B$, and $k_B$ are the Avogadro number, Bohr magneton, and Boltzmann constant, respectively. We obtain $p_{eff}$ = 15.21 $\mu_B$, related to $q_C$ as $p_{eff}^2 = q_C(q_C+2)$, yielding $q_C$ = 14.244 $\mu_B$.  The saturation moment ($q_S$) is calculated from the \textit{M(H)} curve at \textit{T} = 2 K, yielding $q_S$ =  8.246 $\mu_B$. Consequently, $q_C/q_S = 1.727>1$ confirming the itinerant magnetic nature of Co$_{6.5}$Ru$_{1.5}$Zn$_8$Mn$_4$.
To understand the occurrence of the unusual minima at $T_X$ in more detail, we investigate \textit{T} dependence of $M$, for a slow \textit{T} ramp rate of 1 K min$^{-1}$ in the ZFCW mode from 150 - 250 K under various applied \textit{H} [50 Oe to 1000 Oe ($\Delta H_{app}$ = 50 Oe), 1000 Oe to 1500 Oe ($\Delta H_{app}$ = 100 Oe), 2000 Oe, 3000 Oe, and 5000 Oe]  (see supplementary information). Based on Tikhonov regularization\cite{stickel2010data}, a statistical method was employed to remove spurious noise signals appearing in traditional finite difference calculation for numerical derivatives (here, $\frac{dM}{dT}$)\cite{bocarsly2018magnetoentropic}. For the low \textit{H} region ( $H<$ 800 Oe), two minima are clearly observed in $\frac{dM}{dT}$ versus $T$ curves (inset (III) of \text{Fig.} \ref{MCE}(a)), and the corresponding transition temperatures, $T_C$ and $T_X$, gradually shifts towards slightly lower (red dotted line) and higher \textit{T} (green dotted line), respectively, with increasing \textit{H}. On the other hand, for $H$ $>$ 800 Oe, the individual transitions collapse to a single minimum and shift toward higher \textit{T} for growing \textit{H} (black arrow). \text{Figure} \ref{MCE}(b) shows the contour plot of $\textit{T}$ versus \textit{H} phase diagram of d$\chi/dT$, where the color bar represents the magnitude of d$\chi/dT$. In the low field regime (\textit{H} $<$ 400 Oe), we observe a positive d$\chi/dT$, bounded by a zero value region below $T_P$, and d$\chi/dT$ shows a sign change (from positive to negative) with increasing \textit{H}. The lowest value of $d\chi/dT$ is observed at the $T_C$ (black spot) along with a neck in between $T_P$ and $T_C$ (\textit{i.e.}, $T_X$) and with increasing \textit{H}, the two minima at $T_C$ and $T_X$ merge and collapse into a single minimum, indicating the presence of multiple magnetic phases transforming into a field-polarized uniform magnetic phase. To investigate the nature of these complicated multiple magnetic phases, we calculate the change in magnetic entropy under the application of external \textit{H}, commonly referred to as magnetocaloric effect (MCE) \cite{ahmed2022comparative}. The isothermal magnetic entropy change, $\Delta S_m ( T, \Delta H)$ can be expressed as,

\begin{equation}
{\label{ME}
\Delta S_m(T,\Delta H) = \int_0^H \left(\frac{\partial M}{\partial T}\right)_H dH}
\end{equation}

where $H$, $M$ and $T$ are the applied magnetic field, magnetization and temperature, respectively. Figure \ref{MCE}(c) shows the isothermal entropy change for different applied \textit{H} within 150 - 250 K, calculated using Eq. (\ref{ME}). Except for very low \textit{H} ($<$ 400 Oe) below $T_P$, $\Delta S_m$ is negative with a minimum near $T_C$, owing to the reduction of magnetic entropy associated with the alignment of paramagnetic moment towards the applied \textit{H}. Figure \ref{MCE}(d) shows the enlarged plot of $\Delta S_m$ versus \textit{H} ( $<$ 1000 Oe), \textit{i.e.} the regime possessing coexisting multiple complicated magnetic phases, as stated before. Below \textit{T$_P$}, for \textit{H} ($<$ 400 Oe), a pocket (wine) is observed which corresponds to an increase in entropy enveloped by no change of entropy. It is well established that the SkL shows higher entropy than the conical/helical phases which is consistent with the idea of thermal fluctuations assisting in the stabilization of topologically protected skyrmionic lattice\cite{jamaluddin2019robust,bocarsly2018magnetoentropic}. Below $T_P$ under zero applied field we get, $\Delta S_m \approx 0$, associated with the ground conical/helical state which evolves to a small positive value on the application of magnetic field corresponding to the stabilization of SkL\cite{bocarsly2018magnetoentropic}. On further increasing the external field, $\Delta S_m$ changes from positive to negative value (field polarized) via conical/helical ($\Delta S_m \approx 0$). Interestingly, a second MCE response manifesting itself in between $T_C$ and $T_P$ as with two distinct minima is also observed which gradually collapses into a single minima with the increase in applied \textit{H}. This unique behaviour of change in MCE is indicative of a second magnetic phase between the conical/helical and paramagnetic regions. In the next section, we perform critical analysis studies around $T_C$ to understand the nature of magnetic interactions.


\subsection{Magnetic Scaling Analysis}
\begin{figure*}[t]
\includegraphics[width=1\textwidth]{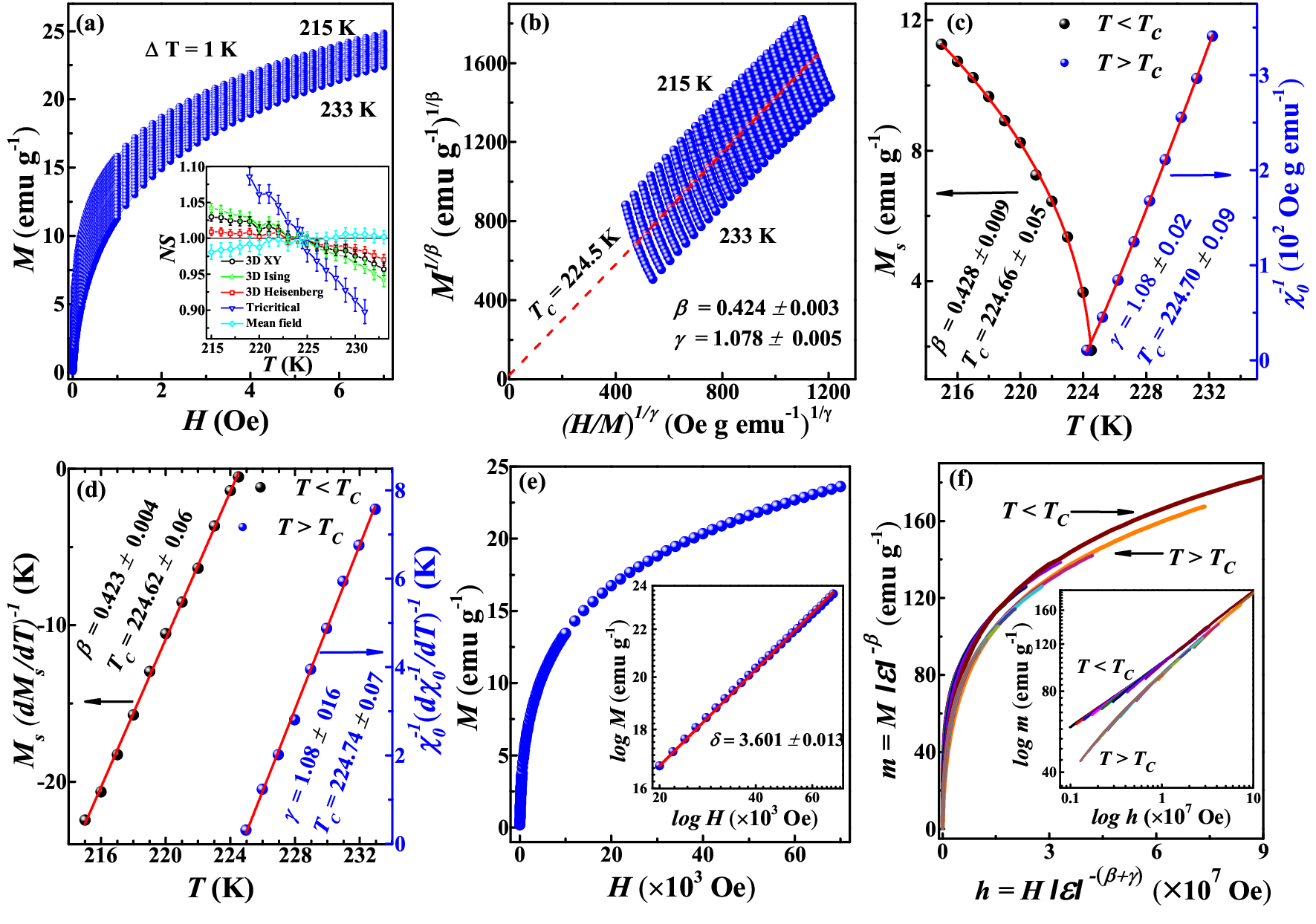}
\centering
\caption[]
{\label{M}(a) Isothermal magnetization curves $M(H)$ at the critical region i.e.  215 K to 233 K, $\Delta T =$ 1 K. Inset shows the variation of normalized slope ($NS$) with temperature for the different theoretical models. (b) Modified Arrot plot (MAP) (c) $M_{s}$(T) and $\chi_0^{-1}(T)$ as a function of temperature. (d) Temperature dependent $M_s (dM_s/dT)^{-1}$ and $\chi_0^{-1}(d\chi_0^{-1}/dT)^{-1}$. (red lines are fitted curves) (e) Critical isotherm curve ($M$ vs. $H$ at 224.5K) inset shows fitted critical isotherm curve on a log-log scale. (f) Variation of renormalized magnetization, $m$ = $|\varepsilon|^{-\beta} M(H,\varepsilon)$ with the renormalized field $h$ = $H|\varepsilon|^{-(\beta+\gamma)}$. (inset shows log-log scale to view clearly at the low field).}
\end{figure*}

According to the scaling theory of continuous second-order magnetic phase transition, the divergence of correlation length near $T_C$, leads to a scaling relation of the saturation moment ($M_S$) and inverse susceptibility ($\chi_0^{-1}$), below and above $T_C$, respectively. The parameters (${\beta}$, ${\gamma}$, ${\delta}$), commonly referred to as critical exponents, are related to the measurable quantities such as magnetization ($M$), susceptibility ($\chi$) and critical magnetization isotherm respectively\cite{sopt1,sopt2} by the following relations,

\begin{equation}
{\label{1}
M_s(T) = M_0 (-\varepsilon)^{-\beta}, \indent \varepsilon < 0, T < T_C,}
\end{equation}
\begin{equation}
{\label{2}
\chi_0^{-1} (T) = \left(\frac{h_0}{M_0}\right) \varepsilon^\gamma,\indent \varepsilon > 0, T > T_C,}
\end{equation}
\begin{equation}
{\label{3}
M = DH^{1/\delta}, \indent \indent  \indent \indent\indent \varepsilon = 0, T = T_C,}
\end{equation}

\indent where $\epsilon = (T - T_C)/T_C$ is the reduced temperature and  $M_0$, $h_0$/$M_0$, and $D$ are critical amplitudes. The critical exponents ${\beta}$, ${\gamma}$, ${\delta}$ are used to define the underlying interaction mechanism of the magnetic system. \text{Figure} \ref{M}(a) shows the magnetic isotherms within the \textit{T} range 215-223 K with a \textit{T} increment of 1 K. Following Banerjee's criterion, a positive slope of the Arrott plot (See Supplementary) would confirm the second-order nature of the phase transition at $T_C$ \cite{Banerjee}. However, the Arrott plot does not produce a set of parallel lines with constant slope indicating the non-viability of the conventional Arrott plot to clarify the magnetic interactions of this system. Hence, we construct the Modified Arrott plot (MAP) such that $\beta$ and $\gamma$ follow the Arrott -Noakes equation, Eqn. \ref{4}, in the asymptotic critical region,\cite{arrotnoakes,theory,theorytcp}

\begin{equation}
{\label{4} \left(\frac{H}{M}\right)^{1/\gamma} = c_1 \left(\frac{T_C-T} {T_C}\right) + c_2 M^{1/\beta}}
\end{equation}

\indent where $c_1$ and $c_2$ are constants, using all the possible universality groups (Landau Mean Field, 3D-XY, 3D-Heisenberg, 3D-Ising,  Tricritical) . All the existing models show a set of quasi-parallel lines. To identify the best model for the spin-interaction of this system we plot the normalized slope, $NS = S(T)/S(T_C)$ \cite{ns},(inset of \text{Fig.} \ref{M}(a)). Our analysis shows that below $T_C$, $NS$ for each isotherm considering the 3D Heisenberg model is closest to the ideal value of $'1'$, however, the scenario changes for $T>T_C$ where Landau Mean Field model dictates the spin interactions. This implies the inability of a single universality class to define the system both before and after the ordering temperature. Following a rigorous iterative method considering the initial values of $\beta$ from 3D Heisenberg and $\gamma$ from Landau Mean Field Model as the critical exponents $\beta$ and $\gamma$ associated with $M_s (T)$ and $\chi_0^{-1} (T)$ below and above $T_C$ respectively, MAP has been constructed. \text{Figure} \ref{M}(b) shows the calculated MAP with the set of $M^{1/0.424}$ versus $(H/M)^{1/1.078}$ parallel lines, and the modified isotherm of $T$ = 224.5 K  passes through the origin, proving the consistency and reliability of the obtained values. The values of $M_s$(T) and $\chi_{0}^{-1}(T)$ are obtained from extrapolating the fitted lines of the isotherms (\text{Fig.} \ref{M}(b)) up to the respective axes. Subsequently, power law fitting of the variation of $M_s$(T) and $\chi_{0}^{-1}(T)$ versus \textit{T} using Eqns. \ref{1} and \ref{2} the values of $\beta$ = 0.428 $\pm$ 0.009; $T_C$  = 224.66 $\pm$ 0.05 and $\gamma$ = 1.08 $\pm$ 0.016; $T_C$ = 224.70 $\pm$ 0.09 are found respectively, shown in \text{Fig.}\ref{M}(c). These values agree well with the ones obtained from the MAP, showing the consistency and intrinsic nature of the obtained parameters. Similar to our previous report\cite{bhattacharya2023critical}, to verify the accuracy of the critical exponents, we also implemented the Kouvel-Fischer analysis on the experimental results\cite{KF}. In this method the plot of $M_S(dM_S/dT)^{-1}$ and $\chi_0^{-1}(d\chi_0^{-1}/dT)^{-1}$ against \textit{T} gives us straight lines with the slope of 1/$\beta$ and 1/$\gamma$ respectively and the intercept at \textit{T} axis giving the $T_C$. Fitting of the experimental results yielded,  $\beta = 0.423 \pm 0.004$; $T_C = 224.62 \pm 0.07$ K and $\gamma = 1.08 \pm 0.016$; $T_C = 224.74 \pm 0.07$ K (\text{Fig.}\ref{M}(d)). The obtained values of the critical exponent from power law fitting and KL analysis are in close agreement indicating the consistency and reliability of the results. We have used the Widom's relation \cite{delta} $\delta = 1+ \frac{\gamma}{\beta}$ for calculating the value of $\delta$ using the value of $\beta$ and $\gamma$ from both the power law and KF-method which yielded $\delta$ = 3.523 $\pm$ 0.008 and $\delta$ = 3.553 $\pm$ 0.009, respectively. To verify the $\delta$ obtained using Widom's criterion, we employ the Eqn.\ref{3} to fit the magnetic isotherm $M(H)$ at $T = T_C =$ 224.5 K in the asymptotic region which yielded $\delta$ = 3.601 $\pm$ 0.013 is shown in \text{Fig.} \ref{M}(e). The value of $\delta$ is in good agreement with the previously obtained values confirming the consistency and reliability of the obtained critical parameters.

Here, the critical exponent $\beta$ = 0.423 $\pm$ 0.004 is in between the Landau mean-field model and the 3D- Heisenberg model and the $\gamma$ = 1.08 $\pm$ 0.016 is close to Landau mean-field model.

Finally, according to scaling law, the magnetic equation in the asymptotic critical region can be expressed as ~\cite{scaling}:
\begin{equation}
{\label{7}
M(H,\varepsilon) = \varepsilon^\beta f_\pm (H\varepsilon^{-(\beta+\gamma)})}
\end{equation}

where, \textit{f}$_+$ and \textit{f}$_-$ represents regular functions at $T>T_C$ and $T<T_C$ respectively. \text{Figure} \ref{M} (f) shows the variation of renormalized magnetization, $m = \varepsilon^{-\beta} M(H,\varepsilon)$ versus renormalized field, $h = H\varepsilon^{-(\beta+\gamma)}$). The inset in \text{Fig.}\ref{M} (f) shows a log-log plot of the scaled magnetic isotherms in the asymptotic field region collapse into two independent universal branches for below and above $T_C$. This scaling analysis verifies that the extracted critical exponents can be reliably used to understand the underlying spin interaction mechanism of this ferromagnetic compound using the experimental magnetic isotherms in the vicinity of $T_C$. With the magnetic interaction of the system getting scaled in two different models above and below $T_C$, the system also provides a great opportunity to verify whether the studied anomalous Hall critical scaling will also reflect the same peculiarity in scaling. In the following section, we scaled the transverse resistivity and examined the above mentioned problem.

\subsection{Hall scaling Analysis}
\begin{figure*}[t]
\includegraphics[width=1\textwidth]{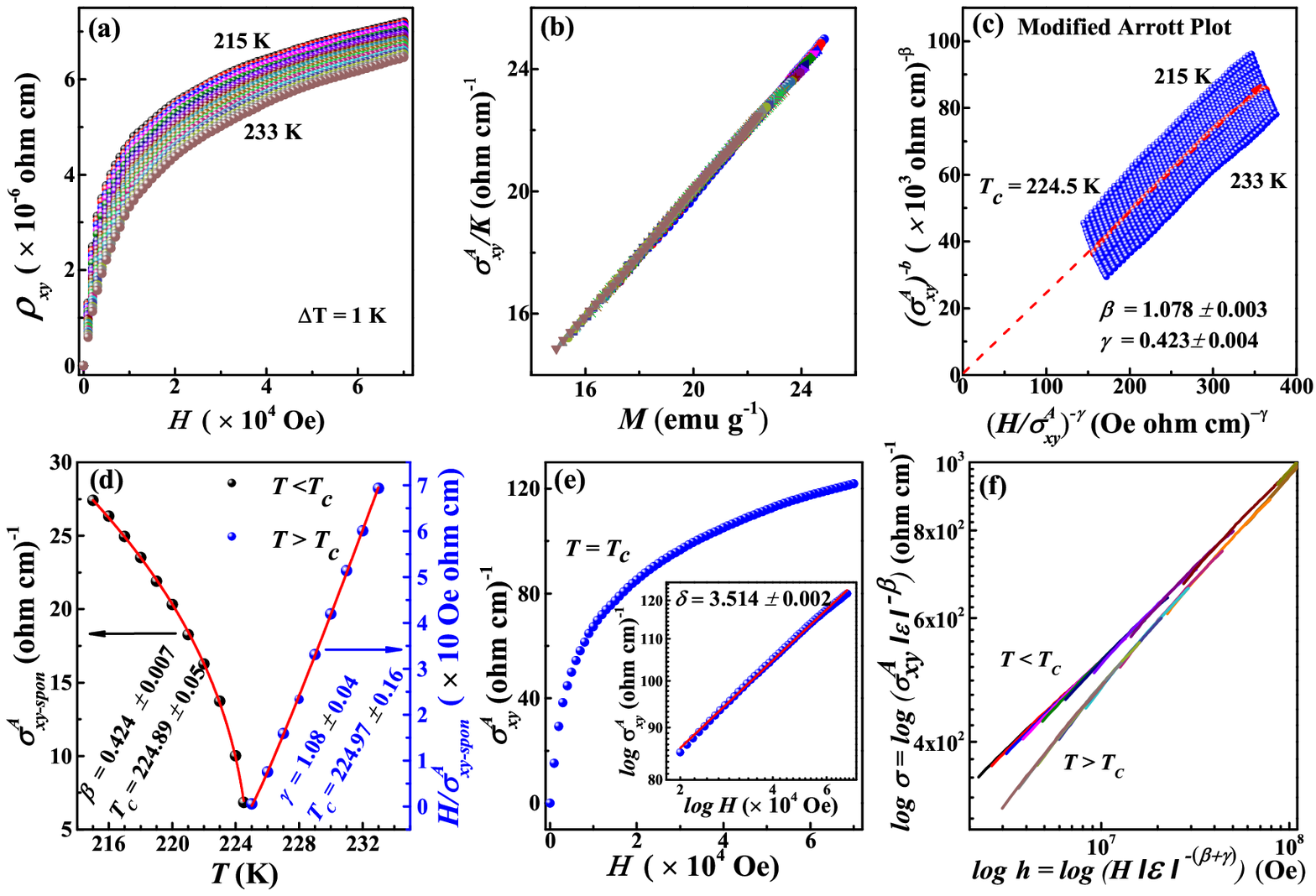}
\centering
\caption[]
{\label{T}(a) Isotherms of Hall resistivity, $\rho_{xy}$ at the critical region 215K to 233K, $\Delta T =$ 1 K (b) Normalized anomalous Hall conductivity (AHC), $\sigma_{xy}^A /K(T)$ vs. magnetization ($M$) for different temperatures (c) Modified Arrot plot (d) Power law fitting: Plot of spontaneous AHC $\sigma_{xy-spon}^A$ and ${H}/{\sigma_{xy}^A})^{1/\gamma}$ with $T$ (red lines are fitted curves gives the value of $\beta$ and $\gamma$ with $T_C$) (e) Critical AHC isotherm, $\sigma_{xy}^A (T_C)$ vs. $H$ (inset shows the fitted curve at high field region which gives the value of $\delta =$ 3.514 $\pm$ 0.002) (f) Scaling plot at log - log scale: renormalized AHC, $\sigma_{xy}^A |\epsilon|^{-\beta}$ vs. renormalized field $H |\varepsilon|^{-(\beta+\gamma)}$ shows clear branching of isotherm above and below $T_C$.}
\end{figure*}

To further investigate the unconventional nature of the occurrence of different universality classes and to confirm its intrinsic character above and below $T_C$, we utilize a recently developed method of critical behaviour analysis using the anomalous Hall conductivity\cite{bhattacharya2023critical}. For a ferro or ferrimagnetic system, the anomalous Hall resistivity can be expressed by the following empirical relation ~\cite{hall1,taguchi2001spin,lee2004dissipationless,paschen2004hall,miyasato2007crossover,fang2003anomalous,tian2009proper}

The recent development of the critical behaviour analysis in ferromagnet using the anomalous Hall conductivity isotherms \cite{bhattacharya2023critical} in the critical regime, provides us with a tool to further verify the validity of the crossover behaviour across the transition temperature. The anomalous Hall resistivity in ferro/ferri magnetic systems can be empirically written as\cite{hall1,taguchi2001spin}, $\rho_{xy} = R_0H+4\pi R_s M$, where $H$ is the external magnetic field perpendicular to the applied current direction, $R_0 = 1/n_c|e|$ is the ordinary Hall coefficient inversely related to electronic charge ($e$) and carrier density ($n_c$) and $R_s$ is the anomalous Hall coefficient. Figure \ref{T}(a) shows the experimental data of transverse Hall resistivity isotherms ranging from 215 - 233 K with an increment $\Delta T$ = 1 K. We evaluate $R_0$ by fitting the experimentally measured total Hall resistivity at field range 5-7 T, yielding a hole-type carrier with density $n_c =$ 2.5 $\pm$ 0.27 $\times 10^{22} cm^{-3}$. According to the theory of Karplus and Luttinger \cite{karplus1954hall}, the anomalous Hall conductivity (AHC) arising from the intrinsic contribution is linearly proportional to the macroscopic magnetism \cite{husmann2006temperature} leading to the following expression for AHC as $\sigma_{xy} = \frac{\rho_{xy}}{\rho_{xy}^2+\rho_{xx}^2} \approx \frac{\rho_{xy}}{\rho_{xx}^2} = \frac{R_0H + 4\pi R_s M}{\rho_{xx}^2} = \sigma_{xy}^N + \sigma_{xy}^A$, where $\sigma_{xy}^N$ and $\sigma_{xy}^A $ are the normal and anomalous Hall conductivity (AHC), respectively, and AHC can be written as $\sigma_{xy}^A = (\sigma_{xy} - \sigma_{xy}^N) = \frac{4\pi R_s M}{\rho_{xx}^2}$. To determine the origin of AHC, we investigated the power-law relation of $R_s$ with $\rho_{xx}$ [i.e. $R_s \propto \rho_{xx}^\alpha$ ], where a linear dependence ($\alpha = 1$) is indicative of dominant extrinsic skew-scattering contribution while $\alpha = $ 2 is attributed to intrinsic and side-jump (SJ) mechanisms. \cite{lavine1961extraordinary,smit1955spontaneous,berger1970side,karplus1954hall,onoda2006intrinsic,zeng2006linear,jungwirth2002anomalous}. A scaling approach is performed to determine the power law dependence of $R_s$ on $\rho_{xx}$ by introducing,  $G(R_s,\rho_{xx})$ =  $4\pi R_s / \rho_{xx}^2$ hence scaling AHC with magnetization ($M$) as $\sigma_{xy}^A$ = $M.G(R_s,\rho_{xx})$\cite{jiang2010scaling} shown in \text{Fig.} \ref{T}(b). Implementing the constraint that the intercept of linear fitting of $\sigma_{xy}^A$ with $M$ in the asymptotic range of $M$ is always zero, i.e., $\sigma_{xy}^A = 0$. The $G(R_s, \rho_{xx})$ is found to be constant over the measured range of $T$ which confirms the quadratic relation between $R_s$ and $\rho_{xx}$ ($R_s = K(T)\rho_{xx}^2$) and hence establishing intrinsic or side-jump mechanism as a dominant contributor in AHC. With the absence of a high Z (atomic number) element providing large SOC, $E_{so}/E_F$ ( $E_{so}$ = spin-orbit interaction energy and $E_F$ = Fermi energy) ratio determining the contribution of SJ-mechanism to AHC becomes negligible.  So, the origin of AHC is attributed to the intrinsic contribution and hence we  rewrite the modified Arrott Nokes equation for Hall conductivity\cite{arrotnoakes} simply  by replacing magnetization (\textit{M}) with $\sigma_{xy}^A$ as,

\begin{figure*}[t]
\includegraphics[width=1.0\textwidth]{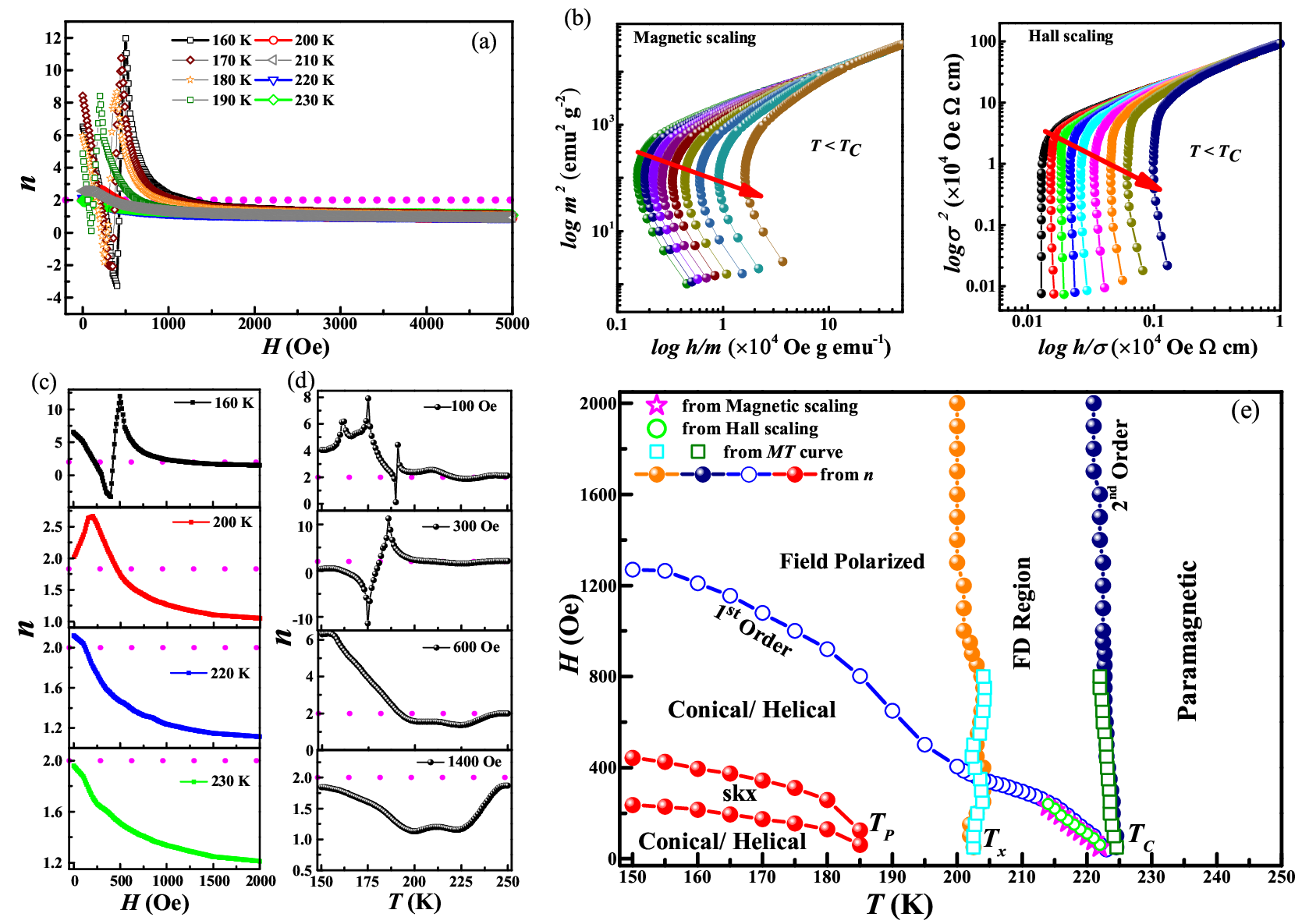}
\centering
\caption[]
{\label{P}(a) $n$ vs. $H$ plot at different temperatures: $T$ = 160 K- 230 K, $\Delta T =$ 10 K. $n$ varies monotonically at high field and diverges from $n$ = 2 (shown by dotted line); (b) $m^2$ vs. $h/m$ and $\sigma^2$ vs. $h/\sigma$ : low field region of the .scaled isotherms for $T<T_C$ in log-log scale. (c) shows zoomed individual $n$ vs. $H$ curves which show multiple crossing across $n$ = 2, signifying transition between phases of first and second order and sudden sign change of n (positive-negative) corresponding to sign change of MCE. (d) Temperature variation of $n$ under different applied fields. The crossover of $n$ vs. $T$ curve across $n$ = 2 signifies the temperature corresponding to phase transition (first- second). The double hump on each $n-T$ curve: corresponds to (PM-FM) and FD phase (e) magnetic phase diagram at low field region in the vicinity of PM-FM transition temperature.}
\end{figure*}

\begin{equation}
{\label{10}
\left(\frac{\sigma_{xy}^A}{\sigma_1}\right)^{1/\beta} = t_1\left(\frac{T_c-T} {T_c}\right)+ t_2\left(\frac{H}{\sigma_{xy}^A}\right)^{1/\gamma} }
\end{equation}

\begin{table*}[ht]
\caption[]
{\label{C} : List of Critical exponents calculated for Co$_{6.5}$Ru$_{1.5}$Zn$_{8}$Mn$_4$ along with five standard theoretical models and some other skyrmion host compounds. (PC = Poly Crystal, SC = Single Crystal, MAP = Modified Arrot Plot).}
\centering
\begin{tabular}{p{3.5 cm} p{2. cm} p{2. cm} p{2. cm} p{2. cm} p{2. cm} p{2. cm}}
\hline
\hline
\noalign{\smallskip}
Composition/ Model   &   References    &   Techniques &       T$_c$ &      $\beta$ &     $\gamma$ &     $\delta$  \\
\noalign{\smallskip}
\hline
\noalign{\smallskip}
& Present work & MAP & 224.5 & 0.424 $\pm$ 0.003 & 1.078$\pm$ 0.005 & 3.601 $\pm$ 0.013\\
Co$_{6.5}$Ru$_{1.5}$Zn$_{8}$Mn$_4$$^{PC}$& ,, & KF & 224.74 $\pm$ 0.07 & 0.423 $\pm$ 0.004 & 1.08 $\pm$ 0.016 & 3.553 $\pm$ 0.009\\
& ,, & Hall& 224.89 $\pm$ 0.05 & 0.424 $\pm$ 0.007 & 1.08 $\pm$ 0.004 & 3.514 $\pm$ 0.002\\
Mean-field & \cite{theory} & - & & 0.5 & 1.0 & 3.0\\
3D-Heisenberg & \cite{theory} & - & & 0.365 & 1.386 & 4.80\\
3D-XY & \cite{theory}& - & & 0.345 & 1.316 & 4.81\\
3D-Ising & \cite{theory} & - & & 0.325 & 1.24 & 4.82\\
Tricritical & \cite{theorytcp} & - & & 0.25 & 1.0 & 5.0\\

Fe$_{0.6}$Co$_{0.4}$Si$^{PC}$ & \cite{FeCoSic}& MAP & 53.72(2) & 0.411(3) & 1.325(6) & 4.223(4)\\
Fe$_{0.8}$Co$_{0.2}$Si$^{PC}$ & \cite{jiang2010scaling} & Hall& 36.0 & 0.371(1) & 1.38(2) & 4.78 (1)\\
MnSi$^{SC}$ &\cite{MnSic} & MAP & 30.5 & 0.242(6) & 0.915(3) & 4.734(6)\\
FeGe$^{PC}$ & \cite{FeGec} & MAP & 283.0 & 0.336(4) & 1.352(3) & 5.267(1)\\
GaV$_4$S$_8^{PC}$ &\cite{GaV4S8c} & MAP& 12.0 & 0.220(2) & 0.909(5) & 5.132(1)\\
\noalign{\smallskip}
\hline
\hline
\end{tabular}
\end{table*}

where $\sigma_1$ is a material specific constant and $t_1$ and $t_2$ are constants. Using the same iterative method as employed for the critical scaling of magnetic isotherms and subsequently starting with values of $\beta$ and $\gamma$ from 3D Heisenberg and Mean Field theory for $T<T_C$ and $T>T_C$, respectively, we obtain $\beta$ = 0.423 $\pm$ 0.004 and $\gamma$ = 1.078 $\pm$ 0.003 for scaled AHC isotherms. Using these obtained values, we constructed the ${\sigma_{xy}^A}^{1/0.423}$ versus $(H/\sigma_{xy}^A)^{1/1.078}$ plot. \text{Figure} \ref{T}(c) shows a set of parallel scaled AHC isotherms in an excellent match with Eqn.\ref{10} with the critical isotherm of $T$ = 224.5 K passing through the origin as indicated by the red dashed line. The intercepts of AHC isotherms below $T_C$ on $\sigma_{xy}^A$ axis at $H$ = 0, yield the spontaneous AHC ($\sigma_{xy-spon}^A$), analogous to the spontaneous magnetization $M_s$ which is governed by power law relation $\sigma_{xy-spon}^A \propto |\varepsilon|^{-\beta}$. While the AHC isotherm intercept on the ordinate (${H}/{\sigma_{xy}^A})^{1/\gamma}$ axis above $T_C$ gives the transport analogue of inverse susceptibility in magnetic scaling, vary with reduced temperature as $\epsilon$ as ${H}/{\sigma_{xy}^A} \propto |\varepsilon|^{\gamma}$. The variation of $\sigma_{xy-spon}^A$ and ${H}/{\sigma_{xy}^A}$ with temperature is shown in \text{Fig.} \ref{T}(d). Considering the proportionality of $\sigma_{xy-spon}^A$ and ${H}/{\sigma_{xy}^A}$ with $\epsilon^{-\beta}$ and $\epsilon^{\gamma}$, is fitted using the same power law relation we utilized in the magnetic isotherm scaling until the self-consistency of the exponents is achieved. This yielded the values of $\beta$ = 0.424 $\pm$ 0.007, $\gamma$ = 1.08 $\pm$ 0.004 which are in excellent agreement with the magnetic scaling analysis. The critical AHC isotherm at $T_C = 224.5$ K is fitted using the relation $\sigma_{xy}^A \propto H^{1/\delta}$ (\text{Fig.} \ref{T}(e)) which gives the value $\delta$ = 3.514 $\pm$ 0.002, further validating the robust and self-consistent nature of the critical parameters. Interestingly, the critical exponent $\gamma$ is close to the mean-field model, while  $\beta$ is in between the mean-field and 3D-Heisenberg model verifying that the system does not belong to a single universality class. With the establishment of linearity relation of $\sigma_{xy}^A$ with $M$ we rewrite the magnetic equation of state (Eqn.\ref{7}) by replacing the magnetization with $\sigma_{xy}^A$ as,

\begin{equation}
{\label{11}
\sigma_{xy}^A(H,\varepsilon)= |\varepsilon|^\beta g_\pm (H\varepsilon^{-(\beta+\gamma)})}
\end{equation}

where $g_+$ and $g_-$ are functions above and below $T_C$ respectively. Figure \ref{T}(f) shows the two separate branching of AHC isotherms on either side of $T_C$, clearly establishing that magnetic and magnetotransport scaling on the same footing to understand the interaction mechanism underlying the precursor effect above $T_C$ leading up to ferromagnetic ordering where an entirely different universality class governs the spin interaction i.e. the crossover behaviour across transition temperature. Before this study, the scaling of AHC for itinerant ferromagnetic materials has been reported for either short-range 3D Heisenberg \cite{jiang2010scaling, bhattacharya2023critical} or long-range Mean Field\cite{jiang2009critical} universality class dictating the spin interaction on either side of $T_C$ making our observation unique for the understanding the AHC scaling behaviour.


\subsection{Spin Interactions and Phase Diagram}

Table.\ref{C} sums up the values of the critical parameters calculated from both magnetization as well as anomalous Hall conductivity isotherms and compares them with existing literature. The coherency of values obtained from both methods in turn proves that crossover behaviour between the universality class of spin interaction across $T_C$ as seen in magnetic critical analysis is also reflected in the critical scaling of anomalous Hall conductivity isotherms. To gain a better insight into the range of interaction we turn towards the renormalization theory which states that, the strength of symmetric interaction ($J(r)$) is a function of the radial distance ($r$) between the interacting spins $\Vec{S_i}$ and $\Vec{S_j}$ and vary as $J(r) \approx r^{-(d+\sigma)}$ \cite{fisher1972critical}, where $d$ is the dimension of the system and $\sigma$ is the range of interaction.

\begin{equation}
\begin{aligned}
\label{13}
\gamma \approx  &  1 + \frac{4(n+2)}{d(n+8)}\Delta \sigma +\frac{8(n+2)(n-4)}{d^2(n+8)^2} \\
& \left[1+\frac{2G\left(\frac{d}{2}\right)(7n+20)}{(n-4)(n+8)}\right]\Delta\sigma^2
\end{aligned}
\end{equation}

where, $\bigtriangleup\sigma$ = $\sigma - d/2$, $G(d/2) = 3 - 1/4(d/2)^2$ and $n$ is the dimensionality of the sample. For bulk materials, by fixing the value of $d$ = 3 and varying the value of $n$ ($n$ = 0, 1, 2, and 3) such that Eqn.\ref{13} gives the value of $\gamma$ and $\delta$ as close to that obtained from Modified Arrott plots of magnetic and Hall isotherm. With the $\sigma$ defining the range of interaction, as $\sigma <$ 2 for short-range interaction and $\sigma \geq$ 2 for long-range interaction following the mean-field model where $J(r)$ varies slower or faster than $r^{-5}$, respectively. For our studied system, starting the iterative method with {d:n}={3:3} as per 3D Heisenberg Model, we arrive at $\sigma$ = 1.617, making the $J(r)$ vary as $r^{-4.617}$. This ranges in between the mean field model and the short-range 3D Heisenberg model implying the inability of a single universality class to define the spin interaction in the critical temperature regime for this system, both above and below the transition temperature.

To further understand the crossover behaviour of spin interaction and identify the different phases co-existing in the critical temperature regime we generated the magnetic phase diagram using the $M(T)$ curves under various applied \textit{H} and the scaled $m$ versus $h$ isotherms.
Figure \ref{MCE}(a) shows the variation of  $dM/dT$ versus $T$ under different applied fields where two different transitions are clearly observed, one corresponding to the PM - FM (second order) at 224.5 K and a second transition at 202.5 K. These two transitions are also reflected in MCE curve shown in \text{Fig.} \ref{MCE}(d). This kind of precursor phase in between the transition from paramagnetic to field polarized states is also observed in low-field region of cubic helimagnets like FeGe\cite{wilhelm2011precursor}. This board phase arises due to the suppression of the increase in correlation length as $T$ approaches $T_C$ due to critical spin fluctuations driving the transition towards continuous crossover and making it a textbook example of 3D Heisenberg transition which is reflected in the critical scaling of both $M(H)$ and $\sigma_{xy}^A$ isotherms\cite{wilhelm2012confinement}. For getting a clear picture of the phase boundary in the low field critical regime, we consider the scaled $m^2$ vs. $h/m$ and $\sigma^2$ vs. $h/\sigma$ curves for $T<T_C$. Figure \ref{P}(d) shows the ungrouping of the scaled magnetic and AHC isotherms at the low field region, interestingly with a change of slope corresponding to the first to second order transition \cite{GaV4S8c, FeCoSic}. We estimated the field values ($H_{k}$) corresponding to the knee in both scaled magnetic and anomalous Hall conductivity from the point of inflexion i.e., $\frac{dm^2} {d(h/m)}=0$ and $\frac{d\sigma^2}{d(h/m)}$ = 0, and used it to construct the phase diagram separating the non-collinear conical/helical from fluctuation mediated state. The $H_{k}$ shows a non-monotonic behaviour defining a continuous phase boundary and it converges excellently from both approaches. To look into the phase evolution arising due to suppression of the critical fluctuations and increase in the correlation length at lower temperatures, we scaled $\Delta S_m$ with $H$ ($\Delta S_m \propto H^n$) for each field and temperature locally \cite{shen2002magnetocaloric}. Figure \ref{P}(a) shows the field variation of the power coefficient $n$, where each individual isotherm shows a monotonic increase as the applied field decreases from the asymptotic limit. According to \textit{Law et al} \cite{law2018quantitative} the scenario changes at $H = H_{cross}$ where the value of $n$ shows a crossover from $n<$ 2 to $n>$ 2 subsequently providing the field value for second to first-order phase transition (\text{Fig.} \ref{P}(b)). This crossover behaviour is also present for a temperature variation of $n$ under different applied fields (\text{Fig.} \ref{P}(c)) which chalks out the phase boundary of FP, Skx and conical/helical phase in the $H-T$ phase diagram. These values of $H_{cross}$ converge excellently with the $H_{k}$ in the critical temperature regime.

\section{Conclusion:}

To sum up, we have studied and established a one-to-one correspondence between the non-trivial critical scaling using anomalous Hall conductivity isotherms and the traditional magnetization-based approach for systems exhibiting a crossover between two different universality classes governing the symmetric spin interaction across the ferromagnetic transition. The obtained self-consistent critical exponents of $\beta = 0.423\pm 0.004$, $\gamma = 1.08 \pm 0.016$ and $\delta = 3.553 \pm 0.009$ from the MAP of AHC and magnetic isotherms along with the radial variation of interaction strength, $J(r) \approx r^{-4.617}$, clearly establishes the short-range 3D Heisenberg and long-range Mean field model as the dictating mechanism of spin interaction below and above $T_C$, respectively. Our experimental results indicate a fluctuation-disordered phase adjacent to the $T_C$ with a low-field skyrmionic pocket embedded in the conical phase. The phase boundary of the SkX pocket has been chalked out by the quantitative criterion analysis of the order of phase transition using the scaling of the magnetocaloric effect. This above understanding of crossover behaviour in the symmetric exchange mechanism across the paramagnetic to ferromagnetic transition from magnetic isotherm scaling and its consistent projection in the scaling of AHC isotherms would catapult a new direction of critical phenomenon investigation using the AHE in similar skyrmion hosting bulk materials, heterostructures and mesoscopic devices.

\section{Acknowledgement:}

A.A. and A.B. would like to acknowledge SINP, India, and the Department of Atomic Energy (DAE), India for their fellowship. The authors will like to acknowledge Prof. Biswarup Satpati, SINP for the TEM EDX measurements.

\bibliography{ref}

\begin{thebibliography}{75}%
\makeatletter
\providecommand \@ifxundefined [1]{%
 \@ifx{#1\undefined}
}%
\providecommand \@ifnum [1]{%
 \ifnum #1\expandafter \@firstoftwo
 \else \expandafter \@secondoftwo
 \fi
}%
\providecommand \@ifx [1]{%
 \ifx #1\expandafter \@firstoftwo
 \else \expandafter \@secondoftwo
 \fi
}%
\providecommand \natexlab [1]{#1}%
\providecommand \enquote  [1]{``#1''}%
\providecommand \bibnamefont  [1]{#1}%
\providecommand \bibfnamefont [1]{#1}%
\providecommand \citenamefont [1]{#1}%
\providecommand \href@noop [0]{\@secondoftwo}%
\providecommand \href [0]{\begingroup \@sanitize@url \@href}%
\providecommand \@href[1]{\@@startlink{#1}\@@href}%
\providecommand \@@href[1]{\endgroup#1\@@endlink}%
\providecommand \@sanitize@url [0]{\catcode `\\12\catcode `\$12\catcode
  `\&12\catcode `\#12\catcode `\^12\catcode `\_12\catcode `\%12\relax}%
\providecommand \@@startlink[1]{}%
\providecommand \@@endlink[0]{}%
\providecommand \url  [0]{\begingroup\@sanitize@url \@url }%
\providecommand \@url [1]{\endgroup\@href {#1}{\urlprefix }}%
\providecommand \urlprefix  [0]{URL }%
\providecommand \Eprint [0]{\href }%
\providecommand \doibase [0]{https://doi.org/}%
\providecommand \selectlanguage [0]{\@gobble}%
\providecommand \bibinfo  [0]{\@secondoftwo}%
\providecommand \bibfield  [0]{\@secondoftwo}%
\providecommand \translation [1]{[#1]}%
\providecommand \BibitemOpen [0]{}%
\providecommand \bibitemStop [0]{}%
\providecommand \bibitemNoStop [0]{.\EOS\space}%
\providecommand \EOS [0]{\spacefactor3000\relax}%
\providecommand \BibitemShut  [1]{\csname bibitem#1\endcsname}%
\let\auto@bib@innerbib\@empty
\bibitem [{\citenamefont {Zhang}\ \emph {et~al.}(2015)\citenamefont {Zhang},
  \citenamefont {Menzel}, \citenamefont {Jin}, \citenamefont {Du},
  \citenamefont {Ge}, \citenamefont {Zhang}, \citenamefont {Pi}, \citenamefont
  {Tian},\ and\ \citenamefont {Zhang}}]{MnSic}%
  \BibitemOpen
  \bibfield  {author} {\bibinfo {author} {\bibfnamefont {L.}~\bibnamefont
  {Zhang}}, \bibinfo {author} {\bibfnamefont {D.}~\bibnamefont {Menzel}},
  \bibinfo {author} {\bibfnamefont {C.}~\bibnamefont {Jin}}, \bibinfo {author}
  {\bibfnamefont {H.}~\bibnamefont {Du}}, \bibinfo {author} {\bibfnamefont
  {M.}~\bibnamefont {Ge}}, \bibinfo {author} {\bibfnamefont {C.}~\bibnamefont
  {Zhang}}, \bibinfo {author} {\bibfnamefont {L.}~\bibnamefont {Pi}}, \bibinfo
  {author} {\bibfnamefont {M.}~\bibnamefont {Tian}},\ and\ \bibinfo {author}
  {\bibfnamefont {Y.}~\bibnamefont {Zhang}},\ }\bibfield  {title} {\bibinfo
  {title} {Critical behavior of the single-crystal helimagnet mnsi},\
  }\href@noop {} {\bibfield  {journal} {\bibinfo  {journal} {Physical Review
  B}\ }\textbf {\bibinfo {volume} {91}},\ \bibinfo {pages} {024403} (\bibinfo
  {year} {2015})}\BibitemShut {NoStop}%
\bibitem [{\citenamefont {Zhang}\ \emph {et~al.}(2016)\citenamefont {Zhang},
  \citenamefont {Han}, \citenamefont {Ge}, \citenamefont {Du}, \citenamefont
  {Jin}, \citenamefont {Wei}, \citenamefont {Fan}, \citenamefont {Zhang},
  \citenamefont {Pi},\ and\ \citenamefont {Zhang}}]{FeGec}%
  \BibitemOpen
  \bibfield  {author} {\bibinfo {author} {\bibfnamefont {L.}~\bibnamefont
  {Zhang}}, \bibinfo {author} {\bibfnamefont {H.}~\bibnamefont {Han}}, \bibinfo
  {author} {\bibfnamefont {M.}~\bibnamefont {Ge}}, \bibinfo {author}
  {\bibfnamefont {H.}~\bibnamefont {Du}}, \bibinfo {author} {\bibfnamefont
  {C.}~\bibnamefont {Jin}}, \bibinfo {author} {\bibfnamefont {W.}~\bibnamefont
  {Wei}}, \bibinfo {author} {\bibfnamefont {J.}~\bibnamefont {Fan}}, \bibinfo
  {author} {\bibfnamefont {C.}~\bibnamefont {Zhang}}, \bibinfo {author}
  {\bibfnamefont {L.}~\bibnamefont {Pi}},\ and\ \bibinfo {author}
  {\bibfnamefont {Y.}~\bibnamefont {Zhang}},\ }\bibfield  {title} {\bibinfo
  {title} {Critical phenomenon of the near room temperature skyrmion material
  fege},\ }\href@noop {} {\bibfield  {journal} {\bibinfo  {journal} {Scientific
  reports}\ }\textbf {\bibinfo {volume} {6}},\ \bibinfo {pages} {1} (\bibinfo
  {year} {2016})}\BibitemShut {NoStop}%
\bibitem [{\citenamefont {Samatham}\ and\ \citenamefont
  {Suresh}(2018)}]{FeCoSic}%
  \BibitemOpen
  \bibfield  {author} {\bibinfo {author} {\bibfnamefont {S.~S.}\ \bibnamefont
  {Samatham}}\ and\ \bibinfo {author} {\bibfnamefont {K.}~\bibnamefont
  {Suresh}},\ }\bibfield  {title} {\bibinfo {title} {Critical exponents and
  universal magnetic behavior of noncentrosymmetric fe0. 6co0. 4si},\
  }\href@noop {} {\bibfield  {journal} {\bibinfo  {journal} {Journal of
  Physics: Condensed Matter}\ }\textbf {\bibinfo {volume} {30}},\ \bibinfo
  {pages} {215802} (\bibinfo {year} {2018})}\BibitemShut {NoStop}%
\bibitem [{\citenamefont {{\v{Z}}ivkovi{\'c}}\ \emph
  {et~al.}(2014)\citenamefont {{\v{Z}}ivkovi{\'c}}, \citenamefont {White},
  \citenamefont {R{\o}nnow}, \citenamefont {Pr{\v{s}}a},\ and\ \citenamefont
  {Berger}}]{Cu2SeO3c}%
  \BibitemOpen
  \bibfield  {author} {\bibinfo {author} {\bibfnamefont {I.}~\bibnamefont
  {{\v{Z}}ivkovi{\'c}}}, \bibinfo {author} {\bibfnamefont {J.}~\bibnamefont
  {White}}, \bibinfo {author} {\bibfnamefont {H.~M.}\ \bibnamefont
  {R{\o}nnow}}, \bibinfo {author} {\bibfnamefont {K.}~\bibnamefont
  {Pr{\v{s}}a}},\ and\ \bibinfo {author} {\bibfnamefont {H.}~\bibnamefont
  {Berger}},\ }\bibfield  {title} {\bibinfo {title} {Critical scaling in the
  cubic helimagnet cu 2 oseo 3},\ }\href@noop {} {\bibfield  {journal}
  {\bibinfo  {journal} {Physical Review B}\ }\textbf {\bibinfo {volume} {89}},\
  \bibinfo {pages} {060401} (\bibinfo {year} {2014})}\BibitemShut {NoStop}%
\bibitem [{\citenamefont {Liu}\ \emph {et~al.}(2020)\citenamefont {Liu},
  \citenamefont {Wang}, \citenamefont {Zou}, \citenamefont {Zhou},
  \citenamefont {Li}, \citenamefont {Xu}, \citenamefont {Zhang}, \citenamefont
  {Xu}, \citenamefont {Tian}, \citenamefont {Du} \emph {et~al.}}]{GaV4S8c}%
  \BibitemOpen
  \bibfield  {author} {\bibinfo {author} {\bibfnamefont {B.}~\bibnamefont
  {Liu}}, \bibinfo {author} {\bibfnamefont {Z.}~\bibnamefont {Wang}}, \bibinfo
  {author} {\bibfnamefont {Y.}~\bibnamefont {Zou}}, \bibinfo {author}
  {\bibfnamefont {S.}~\bibnamefont {Zhou}}, \bibinfo {author} {\bibfnamefont
  {H.}~\bibnamefont {Li}}, \bibinfo {author} {\bibfnamefont {J.}~\bibnamefont
  {Xu}}, \bibinfo {author} {\bibfnamefont {L.}~\bibnamefont {Zhang}}, \bibinfo
  {author} {\bibfnamefont {J.}~\bibnamefont {Xu}}, \bibinfo {author}
  {\bibfnamefont {M.}~\bibnamefont {Tian}}, \bibinfo {author} {\bibfnamefont
  {H.}~\bibnamefont {Du}}, \emph {et~al.},\ }\bibfield  {title} {\bibinfo
  {title} {Field-induced tricritical behavior in the n{\'e}el-type skyrmion
  host gav 4 s 8},\ }\href@noop {} {\bibfield  {journal} {\bibinfo  {journal}
  {Physical Review B}\ }\textbf {\bibinfo {volume} {102}},\ \bibinfo {pages}
  {094431} (\bibinfo {year} {2020})}\BibitemShut {NoStop}%
\bibitem [{\citenamefont {Neubauer}\ \emph {et~al.}(2009)\citenamefont
  {Neubauer}, \citenamefont {Pfleiderer}, \citenamefont {Binz}, \citenamefont
  {Rosch}, \citenamefont {Ritz}, \citenamefont {Niklowitz},\ and\ \citenamefont
  {B{\"o}ni}}]{app1MnSi}%
  \BibitemOpen
  \bibfield  {author} {\bibinfo {author} {\bibfnamefont {A.}~\bibnamefont
  {Neubauer}}, \bibinfo {author} {\bibfnamefont {C.}~\bibnamefont
  {Pfleiderer}}, \bibinfo {author} {\bibfnamefont {B.}~\bibnamefont {Binz}},
  \bibinfo {author} {\bibfnamefont {A.}~\bibnamefont {Rosch}}, \bibinfo
  {author} {\bibfnamefont {R.}~\bibnamefont {Ritz}}, \bibinfo {author}
  {\bibfnamefont {P.}~\bibnamefont {Niklowitz}},\ and\ \bibinfo {author}
  {\bibfnamefont {P.}~\bibnamefont {B{\"o}ni}},\ }\bibfield  {title} {\bibinfo
  {title} {Topological hall effect in the a phase of mnsi},\ }\href@noop {}
  {\bibfield  {journal} {\bibinfo  {journal} {Physical review letters}\
  }\textbf {\bibinfo {volume} {102}},\ \bibinfo {pages} {186602} (\bibinfo
  {year} {2009})}\BibitemShut {NoStop}%
\bibitem [{\citenamefont {Schulz}\ \emph {et~al.}(2012)\citenamefont {Schulz},
  \citenamefont {Ritz}, \citenamefont {Bauer}, \citenamefont {Halder},
  \citenamefont {Wagner}, \citenamefont {Franz}, \citenamefont {Pfleiderer},
  \citenamefont {Everschor}, \citenamefont {Garst},\ and\ \citenamefont
  {Rosch}}]{app2}%
  \BibitemOpen
  \bibfield  {author} {\bibinfo {author} {\bibfnamefont {T.}~\bibnamefont
  {Schulz}}, \bibinfo {author} {\bibfnamefont {R.}~\bibnamefont {Ritz}},
  \bibinfo {author} {\bibfnamefont {A.}~\bibnamefont {Bauer}}, \bibinfo
  {author} {\bibfnamefont {M.}~\bibnamefont {Halder}}, \bibinfo {author}
  {\bibfnamefont {M.}~\bibnamefont {Wagner}}, \bibinfo {author} {\bibfnamefont
  {C.}~\bibnamefont {Franz}}, \bibinfo {author} {\bibfnamefont
  {C.}~\bibnamefont {Pfleiderer}}, \bibinfo {author} {\bibfnamefont
  {K.}~\bibnamefont {Everschor}}, \bibinfo {author} {\bibfnamefont
  {M.}~\bibnamefont {Garst}},\ and\ \bibinfo {author} {\bibfnamefont
  {A.}~\bibnamefont {Rosch}},\ }\bibfield  {title} {\bibinfo {title} {Emergent
  electrodynamics of skyrmions in a chiral magnet},\ }\href@noop {} {\bibfield
  {journal} {\bibinfo  {journal} {Nature Physics}\ }\textbf {\bibinfo {volume}
  {8}},\ \bibinfo {pages} {301} (\bibinfo {year} {2012})}\BibitemShut {NoStop}%
\bibitem [{\citenamefont {Yu}\ \emph {et~al.}(2012)\citenamefont {Yu},
  \citenamefont {Kanazawa}, \citenamefont {Zhang}, \citenamefont {Nagai},
  \citenamefont {Hara}, \citenamefont {Kimoto}, \citenamefont {Matsui},
  \citenamefont {Onose},\ and\ \citenamefont {Tokura}}]{app3}%
  \BibitemOpen
  \bibfield  {author} {\bibinfo {author} {\bibfnamefont {X.}~\bibnamefont
  {Yu}}, \bibinfo {author} {\bibfnamefont {N.}~\bibnamefont {Kanazawa}},
  \bibinfo {author} {\bibfnamefont {W.}~\bibnamefont {Zhang}}, \bibinfo
  {author} {\bibfnamefont {T.}~\bibnamefont {Nagai}}, \bibinfo {author}
  {\bibfnamefont {T.}~\bibnamefont {Hara}}, \bibinfo {author} {\bibfnamefont
  {K.}~\bibnamefont {Kimoto}}, \bibinfo {author} {\bibfnamefont
  {Y.}~\bibnamefont {Matsui}}, \bibinfo {author} {\bibfnamefont
  {Y.}~\bibnamefont {Onose}},\ and\ \bibinfo {author} {\bibfnamefont
  {Y.}~\bibnamefont {Tokura}},\ }\bibfield  {title} {\bibinfo {title} {Skyrmion
  flow near room temperature in an ultralow current density},\ }\href@noop {}
  {\bibfield  {journal} {\bibinfo  {journal} {Nature communications}\ }\textbf
  {\bibinfo {volume} {3}},\ \bibinfo {pages} {1} (\bibinfo {year}
  {2012})}\BibitemShut {NoStop}%
\bibitem [{\citenamefont {Sampaio}\ \emph {et~al.}(2013)\citenamefont
  {Sampaio}, \citenamefont {Cros}, \citenamefont {Rohart}, \citenamefont
  {Thiaville},\ and\ \citenamefont {Fert}}]{app4}%
  \BibitemOpen
  \bibfield  {author} {\bibinfo {author} {\bibfnamefont {J.}~\bibnamefont
  {Sampaio}}, \bibinfo {author} {\bibfnamefont {V.}~\bibnamefont {Cros}},
  \bibinfo {author} {\bibfnamefont {S.}~\bibnamefont {Rohart}}, \bibinfo
  {author} {\bibfnamefont {A.}~\bibnamefont {Thiaville}},\ and\ \bibinfo
  {author} {\bibfnamefont {A.}~\bibnamefont {Fert}},\ }\bibfield  {title}
  {\bibinfo {title} {Nucleation, stability and current-induced motion of
  isolated magnetic skyrmions in nanostructures},\ }\href@noop {} {\bibfield
  {journal} {\bibinfo  {journal} {Nature nanotechnology}\ }\textbf {\bibinfo
  {volume} {8}},\ \bibinfo {pages} {839} (\bibinfo {year} {2013})}\BibitemShut
  {NoStop}%
\bibitem [{\citenamefont {Karube}\ \emph
  {et~al.}(2016{\natexlab{a}})\citenamefont {Karube}, \citenamefont {White},
  \citenamefont {Reynolds}, \citenamefont {Gavilano}, \citenamefont {Oike},
  \citenamefont {Kikkawa}, \citenamefont {Kagawa}, \citenamefont {Tokunaga},
  \citenamefont {R{\o}nnow}, \citenamefont {Tokura} \emph {et~al.}}]{app5}%
  \BibitemOpen
  \bibfield  {author} {\bibinfo {author} {\bibfnamefont {K.}~\bibnamefont
  {Karube}}, \bibinfo {author} {\bibfnamefont {J.}~\bibnamefont {White}},
  \bibinfo {author} {\bibfnamefont {N.}~\bibnamefont {Reynolds}}, \bibinfo
  {author} {\bibfnamefont {J.}~\bibnamefont {Gavilano}}, \bibinfo {author}
  {\bibfnamefont {H.}~\bibnamefont {Oike}}, \bibinfo {author} {\bibfnamefont
  {A.}~\bibnamefont {Kikkawa}}, \bibinfo {author} {\bibfnamefont
  {F.}~\bibnamefont {Kagawa}}, \bibinfo {author} {\bibfnamefont
  {Y.}~\bibnamefont {Tokunaga}}, \bibinfo {author} {\bibfnamefont {H.~M.}\
  \bibnamefont {R{\o}nnow}}, \bibinfo {author} {\bibfnamefont {Y.}~\bibnamefont
  {Tokura}}, \emph {et~al.},\ }\bibfield  {title} {\bibinfo {title} {Robust
  metastable skyrmions and their triangular--square lattice structural
  transition in a high-temperature chiral magnet},\ }\href@noop {} {\bibfield
  {journal} {\bibinfo  {journal} {Nature materials}\ }\textbf {\bibinfo
  {volume} {15}},\ \bibinfo {pages} {1237} (\bibinfo {year}
  {2016}{\natexlab{a}})}\BibitemShut {NoStop}%
\bibitem [{\citenamefont {Muï¿½hlbauer}\ \emph {et~al.}(2009)\citenamefont
  {Muï¿½hlbauer}, \citenamefont {Binz}, \citenamefont {Jonietz},
  \citenamefont {Pfleiderer}, \citenamefont {Rosch}, \citenamefont {Neubauer},
  \citenamefont {Georgii},\ and\ \citenamefont {Boï¿½ni}}]{app6}%
  \BibitemOpen
  \bibfield  {author} {\bibinfo {author} {\bibfnamefont {S.}~\bibnamefont
  {Muï¿½hlbauer}}, \bibinfo {author} {\bibfnamefont {B.}~\bibnamefont
  {Binz}}, \bibinfo {author} {\bibfnamefont {F.}~\bibnamefont {Jonietz}},
  \bibinfo {author} {\bibfnamefont {C.}~\bibnamefont {Pfleiderer}}, \bibinfo
  {author} {\bibfnamefont {A.}~\bibnamefont {Rosch}}, \bibinfo {author}
  {\bibfnamefont {A.}~\bibnamefont {Neubauer}}, \bibinfo {author}
  {\bibfnamefont {R.}~\bibnamefont {Georgii}},\ and\ \bibinfo {author}
  {\bibfnamefont {P.}~\bibnamefont {Boï¿½ni}},\ }\bibfield  {title}
  {\bibinfo {title} {Skyrmion lattice in a chiral magnet},\ }\href@noop {}
  {\bibfield  {journal} {\bibinfo  {journal} {Science}\ }\textbf {\bibinfo
  {volume} {323}},\ \bibinfo {pages} {915} (\bibinfo {year}
  {2009})}\BibitemShut {NoStop}%
\bibitem [{\citenamefont {Seki}\ \emph {et~al.}(2012)\citenamefont {Seki},
  \citenamefont {Yu}, \citenamefont {Ishiwata},\ and\ \citenamefont
  {Tokura}}]{app7cu2seO3}%
  \BibitemOpen
  \bibfield  {author} {\bibinfo {author} {\bibfnamefont {S.}~\bibnamefont
  {Seki}}, \bibinfo {author} {\bibfnamefont {X.}~\bibnamefont {Yu}}, \bibinfo
  {author} {\bibfnamefont {S.}~\bibnamefont {Ishiwata}},\ and\ \bibinfo
  {author} {\bibfnamefont {Y.}~\bibnamefont {Tokura}},\ }\bibfield  {title}
  {\bibinfo {title} {Observation of skyrmions in a multiferroic material},\
  }\href@noop {} {\bibfield  {journal} {\bibinfo  {journal} {Science}\ }\textbf
  {\bibinfo {volume} {336}},\ \bibinfo {pages} {198} (\bibinfo {year}
  {2012})}\BibitemShut {NoStop}%
\bibitem [{\citenamefont {Yambe}\ and\ \citenamefont
  {Hayami}(2021)}]{centsymm}%
  \BibitemOpen
  \bibfield  {author} {\bibinfo {author} {\bibfnamefont {R.}~\bibnamefont
  {Yambe}}\ and\ \bibinfo {author} {\bibfnamefont {S.}~\bibnamefont {Hayami}},\
  }\bibfield  {title} {\bibinfo {title} {Skyrmion crystals in centrosymmetric
  itinerant magnets without horizontal mirror plane},\ }\href@noop {}
  {\bibfield  {journal} {\bibinfo  {journal} {Scientific reports}\ }\textbf
  {\bibinfo {volume} {11}},\ \bibinfo {pages} {1} (\bibinfo {year}
  {2021})}\BibitemShut {NoStop}%
\bibitem [{\citenamefont {Kurumaji}\ \emph {et~al.}(2019)\citenamefont
  {Kurumaji}, \citenamefont {Nakajima}, \citenamefont {Hirschberger},
  \citenamefont {Kikkawa}, \citenamefont {Yamasaki}, \citenamefont {Sagayama},
  \citenamefont {Nakao}, \citenamefont {Taguchi}, \citenamefont {Arima},\ and\
  \citenamefont {Tokura}}]{gd2pdSi3}%
  \BibitemOpen
  \bibfield  {author} {\bibinfo {author} {\bibfnamefont {T.}~\bibnamefont
  {Kurumaji}}, \bibinfo {author} {\bibfnamefont {T.}~\bibnamefont {Nakajima}},
  \bibinfo {author} {\bibfnamefont {M.}~\bibnamefont {Hirschberger}}, \bibinfo
  {author} {\bibfnamefont {A.}~\bibnamefont {Kikkawa}}, \bibinfo {author}
  {\bibfnamefont {Y.}~\bibnamefont {Yamasaki}}, \bibinfo {author}
  {\bibfnamefont {H.}~\bibnamefont {Sagayama}}, \bibinfo {author}
  {\bibfnamefont {H.}~\bibnamefont {Nakao}}, \bibinfo {author} {\bibfnamefont
  {Y.}~\bibnamefont {Taguchi}}, \bibinfo {author} {\bibfnamefont {T.-h.}\
  \bibnamefont {Arima}},\ and\ \bibinfo {author} {\bibfnamefont
  {Y.}~\bibnamefont {Tokura}},\ }\bibfield  {title} {\bibinfo {title} {Skyrmion
  lattice with a giant topological hall effect in a frustrated
  triangular-lattice magnet},\ }\href@noop {} {\bibfield  {journal} {\bibinfo
  {journal} {Science}\ }\textbf {\bibinfo {volume} {365}},\ \bibinfo {pages}
  {914} (\bibinfo {year} {2019})}\BibitemShut {NoStop}%
\bibitem [{\citenamefont {Uchida}\ \emph {et~al.}(2008)\citenamefont {Uchida},
  \citenamefont {Nagaosa}, \citenamefont {He}, \citenamefont {Kaneko},
  \citenamefont {Iguchi}, \citenamefont {Matsui},\ and\ \citenamefont
  {Tokura}}]{FeGe}%
  \BibitemOpen
  \bibfield  {author} {\bibinfo {author} {\bibfnamefont {M.}~\bibnamefont
  {Uchida}}, \bibinfo {author} {\bibfnamefont {N.}~\bibnamefont {Nagaosa}},
  \bibinfo {author} {\bibfnamefont {J.}~\bibnamefont {He}}, \bibinfo {author}
  {\bibfnamefont {Y.}~\bibnamefont {Kaneko}}, \bibinfo {author} {\bibfnamefont
  {S.}~\bibnamefont {Iguchi}}, \bibinfo {author} {\bibfnamefont
  {Y.}~\bibnamefont {Matsui}},\ and\ \bibinfo {author} {\bibfnamefont
  {Y.}~\bibnamefont {Tokura}},\ }\bibfield  {title} {\bibinfo {title}
  {Topological spin textures in the helimagnet fege},\ }\href@noop {}
  {\bibfield  {journal} {\bibinfo  {journal} {Physical Review B}\ }\textbf
  {\bibinfo {volume} {77}},\ \bibinfo {pages} {184402} (\bibinfo {year}
  {2008})}\BibitemShut {NoStop}%
\bibitem [{\citenamefont {M{\"u}nzer}\ \emph {et~al.}(2010)\citenamefont
  {M{\"u}nzer}, \citenamefont {Neubauer}, \citenamefont {Adams}, \citenamefont
  {M{\"u}hlbauer}, \citenamefont {Franz}, \citenamefont {Jonietz},
  \citenamefont {Georgii}, \citenamefont {B{\"o}ni}, \citenamefont {Pedersen},
  \citenamefont {Schmidt} \emph {et~al.}}]{FeCoSi}%
  \BibitemOpen
  \bibfield  {author} {\bibinfo {author} {\bibfnamefont {W.}~\bibnamefont
  {M{\"u}nzer}}, \bibinfo {author} {\bibfnamefont {A.}~\bibnamefont
  {Neubauer}}, \bibinfo {author} {\bibfnamefont {T.}~\bibnamefont {Adams}},
  \bibinfo {author} {\bibfnamefont {S.}~\bibnamefont {M{\"u}hlbauer}}, \bibinfo
  {author} {\bibfnamefont {C.}~\bibnamefont {Franz}}, \bibinfo {author}
  {\bibfnamefont {F.}~\bibnamefont {Jonietz}}, \bibinfo {author} {\bibfnamefont
  {R.}~\bibnamefont {Georgii}}, \bibinfo {author} {\bibfnamefont
  {P.}~\bibnamefont {B{\"o}ni}}, \bibinfo {author} {\bibfnamefont
  {B.}~\bibnamefont {Pedersen}}, \bibinfo {author} {\bibfnamefont
  {M.}~\bibnamefont {Schmidt}}, \emph {et~al.},\ }\bibfield  {title} {\bibinfo
  {title} {Skyrmion lattice in the doped semiconductor fe 1- x co x si},\
  }\href@noop {} {\bibfield  {journal} {\bibinfo  {journal} {Physical Review
  B}\ }\textbf {\bibinfo {volume} {81}},\ \bibinfo {pages} {041203} (\bibinfo
  {year} {2010})}\BibitemShut {NoStop}%
\bibitem [{\citenamefont {Adams}\ \emph {et~al.}(2012)\citenamefont {Adams},
  \citenamefont {Chacon}, \citenamefont {Wagner}, \citenamefont {Bauer},
  \citenamefont {Brandl}, \citenamefont {Pedersen}, \citenamefont {Berger},
  \citenamefont {Lemmens},\ and\ \citenamefont {Pfleiderer}}]{cu2seO3}%
  \BibitemOpen
  \bibfield  {author} {\bibinfo {author} {\bibfnamefont {T.}~\bibnamefont
  {Adams}}, \bibinfo {author} {\bibfnamefont {A.}~\bibnamefont {Chacon}},
  \bibinfo {author} {\bibfnamefont {M.}~\bibnamefont {Wagner}}, \bibinfo
  {author} {\bibfnamefont {A.}~\bibnamefont {Bauer}}, \bibinfo {author}
  {\bibfnamefont {G.}~\bibnamefont {Brandl}}, \bibinfo {author} {\bibfnamefont
  {B.}~\bibnamefont {Pedersen}}, \bibinfo {author} {\bibfnamefont
  {H.}~\bibnamefont {Berger}}, \bibinfo {author} {\bibfnamefont
  {P.}~\bibnamefont {Lemmens}},\ and\ \bibinfo {author} {\bibfnamefont
  {C.}~\bibnamefont {Pfleiderer}},\ }\bibfield  {title} {\bibinfo {title}
  {Long-wavelength helimagnetic order and skyrmion lattice phase in cu 2 oseo
  3},\ }\href@noop {} {\bibfield  {journal} {\bibinfo  {journal} {Physical
  Review Letters}\ }\textbf {\bibinfo {volume} {108}},\ \bibinfo {pages}
  {237204} (\bibinfo {year} {2012})}\BibitemShut {NoStop}%
\bibitem [{\citenamefont {Yu}\ \emph {et~al.}(2011)\citenamefont {Yu},
  \citenamefont {Kanazawa}, \citenamefont {Onose}, \citenamefont {Kimoto},
  \citenamefont {Zhang}, \citenamefont {Ishiwata}, \citenamefont {Matsui},\
  and\ \citenamefont {Tokura}}]{FeGefilm}%
  \BibitemOpen
  \bibfield  {author} {\bibinfo {author} {\bibfnamefont {X.}~\bibnamefont
  {Yu}}, \bibinfo {author} {\bibfnamefont {N.}~\bibnamefont {Kanazawa}},
  \bibinfo {author} {\bibfnamefont {Y.}~\bibnamefont {Onose}}, \bibinfo
  {author} {\bibfnamefont {K.}~\bibnamefont {Kimoto}}, \bibinfo {author}
  {\bibfnamefont {W.}~\bibnamefont {Zhang}}, \bibinfo {author} {\bibfnamefont
  {S.}~\bibnamefont {Ishiwata}}, \bibinfo {author} {\bibfnamefont
  {Y.}~\bibnamefont {Matsui}},\ and\ \bibinfo {author} {\bibfnamefont
  {Y.}~\bibnamefont {Tokura}},\ }\bibfield  {title} {\bibinfo {title} {Near
  room-temperature formation of a skyrmion crystal in thin-films of the
  helimagnet fege},\ }\href@noop {} {\bibfield  {journal} {\bibinfo  {journal}
  {Nature materials}\ }\textbf {\bibinfo {volume} {10}},\ \bibinfo {pages}
  {106} (\bibinfo {year} {2011})}\BibitemShut {NoStop}%
\bibitem [{\citenamefont {Yasui}\ \emph {et~al.}(2020)\citenamefont {Yasui},
  \citenamefont {Butler}, \citenamefont {Khanh}, \citenamefont {Hayami},
  \citenamefont {Nomoto}, \citenamefont {Hanaguri}, \citenamefont {Motome},
  \citenamefont {Arita}, \citenamefont {Arima}, \citenamefont {Tokura} \emph
  {et~al.}}]{gdru2si2}%
  \BibitemOpen
  \bibfield  {author} {\bibinfo {author} {\bibfnamefont {Y.}~\bibnamefont
  {Yasui}}, \bibinfo {author} {\bibfnamefont {C.~J.}\ \bibnamefont {Butler}},
  \bibinfo {author} {\bibfnamefont {N.~D.}\ \bibnamefont {Khanh}}, \bibinfo
  {author} {\bibfnamefont {S.}~\bibnamefont {Hayami}}, \bibinfo {author}
  {\bibfnamefont {T.}~\bibnamefont {Nomoto}}, \bibinfo {author} {\bibfnamefont
  {T.}~\bibnamefont {Hanaguri}}, \bibinfo {author} {\bibfnamefont
  {Y.}~\bibnamefont {Motome}}, \bibinfo {author} {\bibfnamefont
  {R.}~\bibnamefont {Arita}}, \bibinfo {author} {\bibfnamefont {T.-h.}\
  \bibnamefont {Arima}}, \bibinfo {author} {\bibfnamefont {Y.}~\bibnamefont
  {Tokura}}, \emph {et~al.},\ }\bibfield  {title} {\bibinfo {title} {Imaging
  the coupling between itinerant electrons and localised moments in the
  centrosymmetric skyrmion magnet gdru2si2},\ }\href@noop {} {\bibfield
  {journal} {\bibinfo  {journal} {Nature communications}\ }\textbf {\bibinfo
  {volume} {11}},\ \bibinfo {pages} {5925} (\bibinfo {year}
  {2020})}\BibitemShut {NoStop}%
\bibitem [{\citenamefont {Lim}\ \emph {et~al.}(2020)\citenamefont {Lim},
  \citenamefont {Li}, \citenamefont {Huang}, \citenamefont {Chi}, \citenamefont
  {Zhou}, \citenamefont {Zeng}, \citenamefont {Omar}, \citenamefont {Feng},
  \citenamefont {Rusydi}, \citenamefont {Pennycook} \emph
  {et~al.}}]{lim2020emergent}%
  \BibitemOpen
  \bibfield  {author} {\bibinfo {author} {\bibfnamefont {Z.~S.}\ \bibnamefont
  {Lim}}, \bibinfo {author} {\bibfnamefont {C.}~\bibnamefont {Li}}, \bibinfo
  {author} {\bibfnamefont {Z.}~\bibnamefont {Huang}}, \bibinfo {author}
  {\bibfnamefont {X.}~\bibnamefont {Chi}}, \bibinfo {author} {\bibfnamefont
  {J.}~\bibnamefont {Zhou}}, \bibinfo {author} {\bibfnamefont {S.}~\bibnamefont
  {Zeng}}, \bibinfo {author} {\bibfnamefont {G.~J.}\ \bibnamefont {Omar}},
  \bibinfo {author} {\bibfnamefont {Y.~P.}\ \bibnamefont {Feng}}, \bibinfo
  {author} {\bibfnamefont {A.}~\bibnamefont {Rusydi}}, \bibinfo {author}
  {\bibfnamefont {S.~J.}\ \bibnamefont {Pennycook}}, \emph {et~al.},\
  }\bibfield  {title} {\bibinfo {title} {Emergent topological hall effect at a
  charge-transfer interface},\ }\href@noop {} {\bibfield  {journal} {\bibinfo
  {journal} {Small}\ }\textbf {\bibinfo {volume} {16}},\ \bibinfo {pages}
  {2004683} (\bibinfo {year} {2020})}\BibitemShut {NoStop}%
\bibitem [{\citenamefont {Chowdhury}\ \emph {et~al.}(2021)\citenamefont
  {Chowdhury}, \citenamefont {DuttaGupta}, \citenamefont {Patra}, \citenamefont
  {Tretiakov}, \citenamefont {Sharma}, \citenamefont {Fukami}, \citenamefont
  {Ohno},\ and\ \citenamefont {Singh}}]{chowdhury2021unconventional}%
  \BibitemOpen
  \bibfield  {author} {\bibinfo {author} {\bibfnamefont {R.~R.}\ \bibnamefont
  {Chowdhury}}, \bibinfo {author} {\bibfnamefont {S.}~\bibnamefont
  {DuttaGupta}}, \bibinfo {author} {\bibfnamefont {C.}~\bibnamefont {Patra}},
  \bibinfo {author} {\bibfnamefont {O.~A.}\ \bibnamefont {Tretiakov}}, \bibinfo
  {author} {\bibfnamefont {S.}~\bibnamefont {Sharma}}, \bibinfo {author}
  {\bibfnamefont {S.}~\bibnamefont {Fukami}}, \bibinfo {author} {\bibfnamefont
  {H.}~\bibnamefont {Ohno}},\ and\ \bibinfo {author} {\bibfnamefont {R.~P.}\
  \bibnamefont {Singh}},\ }\bibfield  {title} {\bibinfo {title} {Unconventional
  hall effect and its variation with co-doping in van der waals fe3gete2},\
  }\href@noop {} {\bibfield  {journal} {\bibinfo  {journal} {Scientific
  reports}\ }\textbf {\bibinfo {volume} {11}},\ \bibinfo {pages} {14121}
  (\bibinfo {year} {2021})}\BibitemShut {NoStop}%
\bibitem [{\citenamefont {Chowdhury}\ \emph {et~al.}(2022)\citenamefont
  {Chowdhury}, \citenamefont {Patra}, \citenamefont {DuttaGupta}, \citenamefont
  {Satheesh}, \citenamefont {Dan}, \citenamefont {Fukami},\ and\ \citenamefont
  {Singh}}]{chowdhury2022modification}%
  \BibitemOpen
  \bibfield  {author} {\bibinfo {author} {\bibfnamefont {R.~R.}\ \bibnamefont
  {Chowdhury}}, \bibinfo {author} {\bibfnamefont {C.}~\bibnamefont {Patra}},
  \bibinfo {author} {\bibfnamefont {S.}~\bibnamefont {DuttaGupta}}, \bibinfo
  {author} {\bibfnamefont {S.}~\bibnamefont {Satheesh}}, \bibinfo {author}
  {\bibfnamefont {S.}~\bibnamefont {Dan}}, \bibinfo {author} {\bibfnamefont
  {S.}~\bibnamefont {Fukami}},\ and\ \bibinfo {author} {\bibfnamefont {R.~P.}\
  \bibnamefont {Singh}},\ }\bibfield  {title} {\bibinfo {title} {Modification
  of unconventional hall effect with doping at the nonmagnetic site in a
  two-dimensional van der waals ferromagnet},\ }\href@noop {} {\bibfield
  {journal} {\bibinfo  {journal} {Physical Review Materials}\ }\textbf
  {\bibinfo {volume} {6}},\ \bibinfo {pages} {014002} (\bibinfo {year}
  {2022})}\BibitemShut {NoStop}%
\bibitem [{\citenamefont {Tokunaga}\ \emph
  {et~al.}(2015{\natexlab{a}})\citenamefont {Tokunaga}, \citenamefont {Yu},
  \citenamefont {White}, \citenamefont {R{\o}nnow}, \citenamefont {Morikawa},
  \citenamefont {Taguchi},\ and\ \citenamefont {Tokura}}]{coznmn}%
  \BibitemOpen
  \bibfield  {author} {\bibinfo {author} {\bibfnamefont {Y.}~\bibnamefont
  {Tokunaga}}, \bibinfo {author} {\bibfnamefont {X.}~\bibnamefont {Yu}},
  \bibinfo {author} {\bibfnamefont {J.}~\bibnamefont {White}}, \bibinfo
  {author} {\bibfnamefont {H.~M.}\ \bibnamefont {R{\o}nnow}}, \bibinfo {author}
  {\bibfnamefont {D.}~\bibnamefont {Morikawa}}, \bibinfo {author}
  {\bibfnamefont {Y.}~\bibnamefont {Taguchi}},\ and\ \bibinfo {author}
  {\bibfnamefont {Y.}~\bibnamefont {Tokura}},\ }\bibfield  {title} {\bibinfo
  {title} {A new class of chiral materials hosting magnetic skyrmions beyond
  room temperature},\ }\href@noop {} {\bibfield  {journal} {\bibinfo  {journal}
  {Nature communications}\ }\textbf {\bibinfo {volume} {6}},\ \bibinfo {pages}
  {1} (\bibinfo {year} {2015}{\natexlab{a}})}\BibitemShut {NoStop}%
\bibitem [{\citenamefont {Karube}\ \emph
  {et~al.}(2016{\natexlab{b}})\citenamefont {Karube}, \citenamefont {White},
  \citenamefont {Reynolds}, \citenamefont {Gavilano}, \citenamefont {Oike},
  \citenamefont {Kikkawa}, \citenamefont {Kagawa}, \citenamefont {Tokunaga},
  \citenamefont {R{\o}nnow}, \citenamefont {Tokura} \emph
  {et~al.}}]{karube2016robust}%
  \BibitemOpen
  \bibfield  {author} {\bibinfo {author} {\bibfnamefont {K.}~\bibnamefont
  {Karube}}, \bibinfo {author} {\bibfnamefont {J.}~\bibnamefont {White}},
  \bibinfo {author} {\bibfnamefont {N.}~\bibnamefont {Reynolds}}, \bibinfo
  {author} {\bibfnamefont {J.}~\bibnamefont {Gavilano}}, \bibinfo {author}
  {\bibfnamefont {H.}~\bibnamefont {Oike}}, \bibinfo {author} {\bibfnamefont
  {A.}~\bibnamefont {Kikkawa}}, \bibinfo {author} {\bibfnamefont
  {F.}~\bibnamefont {Kagawa}}, \bibinfo {author} {\bibfnamefont
  {Y.}~\bibnamefont {Tokunaga}}, \bibinfo {author} {\bibfnamefont {H.~M.}\
  \bibnamefont {R{\o}nnow}}, \bibinfo {author} {\bibfnamefont {Y.}~\bibnamefont
  {Tokura}}, \emph {et~al.},\ }\bibfield  {title} {\bibinfo {title} {Robust
  metastable skyrmions and their triangular--square lattice structural
  transition in a high-temperature chiral magnet},\ }\href@noop {} {\bibfield
  {journal} {\bibinfo  {journal} {Nature materials}\ }\textbf {\bibinfo
  {volume} {15}},\ \bibinfo {pages} {1237} (\bibinfo {year}
  {2016}{\natexlab{b}})}\BibitemShut {NoStop}%
\bibitem [{\citenamefont {Tokunaga}\ \emph
  {et~al.}(2015{\natexlab{b}})\citenamefont {Tokunaga}, \citenamefont {Yu},
  \citenamefont {White}, \citenamefont {R{\o}nnow}, \citenamefont {Morikawa},
  \citenamefont {Taguchi},\ and\ \citenamefont {Tokura}}]{tokunaga2015new}%
  \BibitemOpen
  \bibfield  {author} {\bibinfo {author} {\bibfnamefont {Y.}~\bibnamefont
  {Tokunaga}}, \bibinfo {author} {\bibfnamefont {X.}~\bibnamefont {Yu}},
  \bibinfo {author} {\bibfnamefont {J.}~\bibnamefont {White}}, \bibinfo
  {author} {\bibfnamefont {H.~M.}\ \bibnamefont {R{\o}nnow}}, \bibinfo {author}
  {\bibfnamefont {D.}~\bibnamefont {Morikawa}}, \bibinfo {author}
  {\bibfnamefont {Y.}~\bibnamefont {Taguchi}},\ and\ \bibinfo {author}
  {\bibfnamefont {Y.}~\bibnamefont {Tokura}},\ }\bibfield  {title} {\bibinfo
  {title} {A new class of chiral materials hosting magnetic skyrmions beyond
  room temperature},\ }\href@noop {} {\bibfield  {journal} {\bibinfo  {journal}
  {Nature communications}\ }\textbf {\bibinfo {volume} {6}},\ \bibinfo {pages}
  {7638} (\bibinfo {year} {2015}{\natexlab{b}})}\BibitemShut {NoStop}%
\bibitem [{\citenamefont {Karube}\ \emph {et~al.}(2020)\citenamefont {Karube},
  \citenamefont {White}, \citenamefont {Ukleev}, \citenamefont {Dewhurst},
  \citenamefont {Cubitt}, \citenamefont {Kikkawa}, \citenamefont {Tokunaga},
  \citenamefont {R{\o}nnow}, \citenamefont {Tokura},\ and\ \citenamefont
  {Taguchi}}]{karubecoznmn}%
  \BibitemOpen
  \bibfield  {author} {\bibinfo {author} {\bibfnamefont {K.}~\bibnamefont
  {Karube}}, \bibinfo {author} {\bibfnamefont {J.}~\bibnamefont {White}},
  \bibinfo {author} {\bibfnamefont {V.}~\bibnamefont {Ukleev}}, \bibinfo
  {author} {\bibfnamefont {C.}~\bibnamefont {Dewhurst}}, \bibinfo {author}
  {\bibfnamefont {R.}~\bibnamefont {Cubitt}}, \bibinfo {author} {\bibfnamefont
  {A.}~\bibnamefont {Kikkawa}}, \bibinfo {author} {\bibfnamefont
  {Y.}~\bibnamefont {Tokunaga}}, \bibinfo {author} {\bibfnamefont
  {H.}~\bibnamefont {R{\o}nnow}}, \bibinfo {author} {\bibfnamefont
  {Y.}~\bibnamefont {Tokura}},\ and\ \bibinfo {author} {\bibfnamefont
  {Y.}~\bibnamefont {Taguchi}},\ }\bibfield  {title} {\bibinfo {title}
  {Metastable skyrmion lattices governed by magnetic disorder and anisotropy in
  $\beta$-mn-type chiral magnets},\ }\href@noop {} {\bibfield  {journal}
  {\bibinfo  {journal} {Physical Review B}\ }\textbf {\bibinfo {volume}
  {102}},\ \bibinfo {pages} {064408} (\bibinfo {year} {2020})}\BibitemShut
  {NoStop}%
\bibitem [{\citenamefont {Yu}\ \emph {et~al.}(2018)\citenamefont {Yu},
  \citenamefont {Koshibae}, \citenamefont {Tokunaga}, \citenamefont {Shibata},
  \citenamefont {Taguchi}, \citenamefont {Nagaosa},\ and\ \citenamefont
  {Tokura}}]{yu2018transformation}%
  \BibitemOpen
  \bibfield  {author} {\bibinfo {author} {\bibfnamefont {X.}~\bibnamefont
  {Yu}}, \bibinfo {author} {\bibfnamefont {W.}~\bibnamefont {Koshibae}},
  \bibinfo {author} {\bibfnamefont {Y.}~\bibnamefont {Tokunaga}}, \bibinfo
  {author} {\bibfnamefont {K.}~\bibnamefont {Shibata}}, \bibinfo {author}
  {\bibfnamefont {Y.}~\bibnamefont {Taguchi}}, \bibinfo {author} {\bibfnamefont
  {N.}~\bibnamefont {Nagaosa}},\ and\ \bibinfo {author} {\bibfnamefont
  {Y.}~\bibnamefont {Tokura}},\ }\bibfield  {title} {\bibinfo {title}
  {Transformation between meron and skyrmion topological spin textures in a
  chiral magnet},\ }\href@noop {} {\bibfield  {journal} {\bibinfo  {journal}
  {Nature}\ }\textbf {\bibinfo {volume} {564}},\ \bibinfo {pages} {95}
  (\bibinfo {year} {2018})}\BibitemShut {NoStop}%
\bibitem [{\citenamefont {Morikawa}\ \emph {et~al.}(2017)\citenamefont
  {Morikawa}, \citenamefont {Yu}, \citenamefont {Karube}, \citenamefont
  {Tokunaga}, \citenamefont {Taguchi}, \citenamefont {Arima},\ and\
  \citenamefont {Tokura}}]{morikawa2017deformation}%
  \BibitemOpen
  \bibfield  {author} {\bibinfo {author} {\bibfnamefont {D.}~\bibnamefont
  {Morikawa}}, \bibinfo {author} {\bibfnamefont {X.}~\bibnamefont {Yu}},
  \bibinfo {author} {\bibfnamefont {K.}~\bibnamefont {Karube}}, \bibinfo
  {author} {\bibfnamefont {Y.}~\bibnamefont {Tokunaga}}, \bibinfo {author}
  {\bibfnamefont {Y.}~\bibnamefont {Taguchi}}, \bibinfo {author} {\bibfnamefont
  {T.-h.}\ \bibnamefont {Arima}},\ and\ \bibinfo {author} {\bibfnamefont
  {Y.}~\bibnamefont {Tokura}},\ }\bibfield  {title} {\bibinfo {title}
  {Deformation of topologically-protected supercooled skyrmions in a thin plate
  of chiral magnet co8zn8mn4},\ }\href@noop {} {\bibfield  {journal} {\bibinfo
  {journal} {Nano letters}\ }\textbf {\bibinfo {volume} {17}},\ \bibinfo
  {pages} {1637} (\bibinfo {year} {2017})}\BibitemShut {NoStop}%
\bibitem [{\citenamefont {Menzel}\ \emph {et~al.}(2019)\citenamefont {Menzel},
  \citenamefont {Baabe}, \citenamefont {Litterst}, \citenamefont {Steinki},
  \citenamefont {Dietze}, \citenamefont {Sach}, \citenamefont {Rubrecht},
  \citenamefont {S{\"u}llow},\ and\ \citenamefont {Hoser}}]{menzel2019local}%
  \BibitemOpen
  \bibfield  {author} {\bibinfo {author} {\bibfnamefont {D.}~\bibnamefont
  {Menzel}}, \bibinfo {author} {\bibfnamefont {D.}~\bibnamefont {Baabe}},
  \bibinfo {author} {\bibfnamefont {F.}~\bibnamefont {Litterst}}, \bibinfo
  {author} {\bibfnamefont {N.}~\bibnamefont {Steinki}}, \bibinfo {author}
  {\bibfnamefont {K.}~\bibnamefont {Dietze}}, \bibinfo {author} {\bibfnamefont
  {M.}~\bibnamefont {Sach}}, \bibinfo {author} {\bibfnamefont {B.}~\bibnamefont
  {Rubrecht}}, \bibinfo {author} {\bibfnamefont {S.}~\bibnamefont
  {S{\"u}llow}},\ and\ \bibinfo {author} {\bibfnamefont {A.}~\bibnamefont
  {Hoser}},\ }\bibfield  {title} {\bibinfo {title} {Local structure
  determination in helimagnetic},\ }\href@noop {} {\bibfield  {journal}
  {\bibinfo  {journal} {Journal of Physics Communications}\ }\textbf {\bibinfo
  {volume} {3}},\ \bibinfo {pages} {025001} (\bibinfo {year}
  {2019})}\BibitemShut {NoStop}%
\bibitem [{\citenamefont {Bhattacharya}\ \emph {et~al.}(2023)\citenamefont
  {Bhattacharya}, \citenamefont {Ahmed}, \citenamefont {DuttaGupta},\ and\
  \citenamefont {Das}}]{bhattacharya2023critical}%
  \BibitemOpen
  \bibfield  {author} {\bibinfo {author} {\bibfnamefont {A.}~\bibnamefont
  {Bhattacharya}}, \bibinfo {author} {\bibfnamefont {A.}~\bibnamefont {Ahmed}},
  \bibinfo {author} {\bibfnamefont {S.}~\bibnamefont {DuttaGupta}},\ and\
  \bibinfo {author} {\bibfnamefont {I.}~\bibnamefont {Das}},\ }\bibfield
  {title} {\bibinfo {title} {Critical behavior and phase diagram of
  skyrmion-hosting material co3. 6fe4. 4zn8mn4 probed by anomalous hall
  effect},\ }\href@noop {} {\bibfield  {journal} {\bibinfo  {journal} {Journal
  of Alloys and Compounds}\ ,\ \bibinfo {pages} {170274}} (\bibinfo {year}
  {2023})}\BibitemShut {NoStop}%
\bibitem [{\citenamefont {Karube}\ \emph {et~al.}(2018)\citenamefont {Karube},
  \citenamefont {Shibata}, \citenamefont {White}, \citenamefont {Koretsune},
  \citenamefont {Yu}, \citenamefont {Tokunaga}, \citenamefont {R{\o}nnow},
  \citenamefont {Arita}, \citenamefont {Arima}, \citenamefont {Tokura} \emph
  {et~al.}}]{Fecoznmn}%
  \BibitemOpen
  \bibfield  {author} {\bibinfo {author} {\bibfnamefont {K.}~\bibnamefont
  {Karube}}, \bibinfo {author} {\bibfnamefont {K.}~\bibnamefont {Shibata}},
  \bibinfo {author} {\bibfnamefont {J.}~\bibnamefont {White}}, \bibinfo
  {author} {\bibfnamefont {T.}~\bibnamefont {Koretsune}}, \bibinfo {author}
  {\bibfnamefont {X.}~\bibnamefont {Yu}}, \bibinfo {author} {\bibfnamefont
  {Y.}~\bibnamefont {Tokunaga}}, \bibinfo {author} {\bibfnamefont
  {H.}~\bibnamefont {R{\o}nnow}}, \bibinfo {author} {\bibfnamefont
  {R.}~\bibnamefont {Arita}}, \bibinfo {author} {\bibfnamefont
  {T.}~\bibnamefont {Arima}}, \bibinfo {author} {\bibfnamefont
  {Y.}~\bibnamefont {Tokura}}, \emph {et~al.},\ }\bibfield  {title} {\bibinfo
  {title} {Controlling the helicity of magnetic skyrmions in a $\beta$-mn-type
  high-temperature chiral magnet},\ }\href@noop {} {\bibfield  {journal}
  {\bibinfo  {journal} {Physical Review B}\ }\textbf {\bibinfo {volume} {98}},\
  \bibinfo {pages} {155120} (\bibinfo {year} {2018})}\BibitemShut {NoStop}%
\bibitem [{\citenamefont {Pramanik}\ and\ \citenamefont
  {Banerjee}(2009)}]{pramanik2009critical}%
  \BibitemOpen
  \bibfield  {author} {\bibinfo {author} {\bibfnamefont {A.}~\bibnamefont
  {Pramanik}}\ and\ \bibinfo {author} {\bibfnamefont {A.}~\bibnamefont
  {Banerjee}},\ }\bibfield  {title} {\bibinfo {title} {Critical behavior at
  paramagnetic to ferromagnetic phase transition in pr 0.5 sr 0.5 mno 3: A bulk
  magnetization study},\ }\href@noop {} {\bibfield  {journal} {\bibinfo
  {journal} {Physical Review B}\ }\textbf {\bibinfo {volume} {79}},\ \bibinfo
  {pages} {214426} (\bibinfo {year} {2009})}\BibitemShut {NoStop}%
\bibitem [{\citenamefont {Taran}\ \emph {et~al.}(2005)\citenamefont {Taran},
  \citenamefont {Chaudhuri}, \citenamefont {Chatterjee}, \citenamefont {Yang},
  \citenamefont {Neeleshwar},\ and\ \citenamefont {Chen}}]{taran2005critical}%
  \BibitemOpen
  \bibfield  {author} {\bibinfo {author} {\bibfnamefont {S.}~\bibnamefont
  {Taran}}, \bibinfo {author} {\bibfnamefont {B.}~\bibnamefont {Chaudhuri}},
  \bibinfo {author} {\bibfnamefont {S.}~\bibnamefont {Chatterjee}}, \bibinfo
  {author} {\bibfnamefont {H.}~\bibnamefont {Yang}}, \bibinfo {author}
  {\bibfnamefont {S.}~\bibnamefont {Neeleshwar}},\ and\ \bibinfo {author}
  {\bibfnamefont {Y.}~\bibnamefont {Chen}},\ }\bibfield  {title} {\bibinfo
  {title} {Critical exponents of the la0. 7sr0. 3mno3, la0. 7ca0. 3mno3, and
  pr0. 7ca0. 3mno3 systems showing correlation between transport and magnetic
  properties},\ }\href@noop {} {\bibfield  {journal} {\bibinfo  {journal}
  {Journal of applied physics}\ }\textbf {\bibinfo {volume} {98}} (\bibinfo
  {year} {2005})}\BibitemShut {NoStop}%
\bibitem [{\citenamefont {Poon}\ and\ \citenamefont
  {Durand}(1977)}]{poon1977critical}%
  \BibitemOpen
  \bibfield  {author} {\bibinfo {author} {\bibfnamefont {S.}~\bibnamefont
  {Poon}}\ and\ \bibinfo {author} {\bibfnamefont {J.}~\bibnamefont {Durand}},\
  }\bibfield  {title} {\bibinfo {title} {Critical phenomena and magnetic
  properties of an amorphous ferromagnet: Gadolinium-gold},\ }\href@noop {}
  {\bibfield  {journal} {\bibinfo  {journal} {Physical Review B}\ }\textbf
  {\bibinfo {volume} {16}},\ \bibinfo {pages} {316} (\bibinfo {year}
  {1977})}\BibitemShut {NoStop}%
\bibitem [{\citenamefont {Shahi}\ \emph {et~al.}(2014)\citenamefont {Shahi},
  \citenamefont {Singh}, \citenamefont {Kumar}, \citenamefont {Shukla},
  \citenamefont {Ghosh}, \citenamefont {Yadav}, \citenamefont {Nigam},\ and\
  \citenamefont {Chatterjee}}]{MnV2O4}%
  \BibitemOpen
  \bibfield  {author} {\bibinfo {author} {\bibfnamefont {P.}~\bibnamefont
  {Shahi}}, \bibinfo {author} {\bibfnamefont {H.}~\bibnamefont {Singh}},
  \bibinfo {author} {\bibfnamefont {A.}~\bibnamefont {Kumar}}, \bibinfo
  {author} {\bibfnamefont {K.~K.}\ \bibnamefont {Shukla}}, \bibinfo {author}
  {\bibfnamefont {A.~K.}\ \bibnamefont {Ghosh}}, \bibinfo {author}
  {\bibfnamefont {A.~K.}\ \bibnamefont {Yadav}}, \bibinfo {author}
  {\bibfnamefont {A.~K.}\ \bibnamefont {Nigam}},\ and\ \bibinfo {author}
  {\bibfnamefont {S.}~\bibnamefont {Chatterjee}},\ }\bibfield  {title}
  {\bibinfo {title} {{Effect of Zn doping on the magneto-caloric effect and
  critical constants of Mott insulator MnV2O4}},\ }\bibfield  {journal}
  {\bibinfo  {journal} {AIP Advances}\ }\textbf {\bibinfo {volume} {4}},\ \href
  {https://doi.org/10.1063/1.4896955} {10.1063/1.4896955} (\bibinfo {year}
  {2014}),\ \bibinfo {note} {097137},\ \Eprint
  {https://arxiv.org/abs/https://pubs.aip.org/aip/adv/article-pdf/doi/10.1063/1.4896955/12837761/097137\_1\_online.pdf}
  {https://pubs.aip.org/aip/adv/article-pdf/doi/10.1063/1.4896955/12837761/097137\_1\_online.pdf}
  \BibitemShut {NoStop}%
\bibitem [{\citenamefont {Law}\ \emph {et~al.}(2018)\citenamefont {Law},
  \citenamefont {Franco}, \citenamefont {Moreno-Ram{\'\i}rez}, \citenamefont
  {Conde}, \citenamefont {Karpenkov}, \citenamefont {Radulov}, \citenamefont
  {Skokov},\ and\ \citenamefont {Gutfleisch}}]{law2018quantitative}%
  \BibitemOpen
  \bibfield  {author} {\bibinfo {author} {\bibfnamefont {J.~Y.}\ \bibnamefont
  {Law}}, \bibinfo {author} {\bibfnamefont {V.}~\bibnamefont {Franco}},
  \bibinfo {author} {\bibfnamefont {L.~M.}\ \bibnamefont
  {Moreno-Ram{\'\i}rez}}, \bibinfo {author} {\bibfnamefont {A.}~\bibnamefont
  {Conde}}, \bibinfo {author} {\bibfnamefont {D.~Y.}\ \bibnamefont
  {Karpenkov}}, \bibinfo {author} {\bibfnamefont {I.}~\bibnamefont {Radulov}},
  \bibinfo {author} {\bibfnamefont {K.~P.}\ \bibnamefont {Skokov}},\ and\
  \bibinfo {author} {\bibfnamefont {O.}~\bibnamefont {Gutfleisch}},\ }\bibfield
   {title} {\bibinfo {title} {A quantitative criterion for determining the
  order of magnetic phase transitions using the magnetocaloric effect},\
  }\href@noop {} {\bibfield  {journal} {\bibinfo  {journal} {Nature
  communications}\ }\textbf {\bibinfo {volume} {9}},\ \bibinfo {pages} {1}
  (\bibinfo {year} {2018})}\BibitemShut {NoStop}%
\bibitem [{\citenamefont {Xie}\ \emph {et~al.}(2013)\citenamefont {Xie},
  \citenamefont {Thimmaiah}, \citenamefont {Lamsal}, \citenamefont {Liu},
  \citenamefont {Heitmann}, \citenamefont {Quirinale}, \citenamefont {Goldman},
  \citenamefont {Pecharsky},\ and\ \citenamefont {Miller}}]{xie2013beta}%
  \BibitemOpen
  \bibfield  {author} {\bibinfo {author} {\bibfnamefont {W.}~\bibnamefont
  {Xie}}, \bibinfo {author} {\bibfnamefont {S.}~\bibnamefont {Thimmaiah}},
  \bibinfo {author} {\bibfnamefont {J.}~\bibnamefont {Lamsal}}, \bibinfo
  {author} {\bibfnamefont {J.}~\bibnamefont {Liu}}, \bibinfo {author}
  {\bibfnamefont {T.~W.}\ \bibnamefont {Heitmann}}, \bibinfo {author}
  {\bibfnamefont {D.}~\bibnamefont {Quirinale}}, \bibinfo {author}
  {\bibfnamefont {A.~I.}\ \bibnamefont {Goldman}}, \bibinfo {author}
  {\bibfnamefont {V.}~\bibnamefont {Pecharsky}},\ and\ \bibinfo {author}
  {\bibfnamefont {G.~J.}\ \bibnamefont {Miller}},\ }\bibfield  {title}
  {\bibinfo {title} {$\beta$-mn-type co8+ x zn12--x as a defect cubic laves
  phase: Site preferences, magnetism, and electronic structure},\ }\href@noop
  {} {\bibfield  {journal} {\bibinfo  {journal} {Inorganic chemistry}\ }\textbf
  {\bibinfo {volume} {52}},\ \bibinfo {pages} {9399} (\bibinfo {year}
  {2013})}\BibitemShut {NoStop}%
\bibitem [{\citenamefont {Henderson}\ \emph {et~al.}(2021)\citenamefont
  {Henderson}, \citenamefont {Beare}, \citenamefont {Sharma}, \citenamefont
  {Bleuel}, \citenamefont {Clancy}, \citenamefont {Cory}, \citenamefont
  {Huber}, \citenamefont {Marjerrison}, \citenamefont {Pula}, \citenamefont
  {Sarenac} \emph {et~al.}}]{characterization}%
  \BibitemOpen
  \bibfield  {author} {\bibinfo {author} {\bibfnamefont {M.~E.}\ \bibnamefont
  {Henderson}}, \bibinfo {author} {\bibfnamefont {J.}~\bibnamefont {Beare}},
  \bibinfo {author} {\bibfnamefont {S.}~\bibnamefont {Sharma}}, \bibinfo
  {author} {\bibfnamefont {M.}~\bibnamefont {Bleuel}}, \bibinfo {author}
  {\bibfnamefont {P.}~\bibnamefont {Clancy}}, \bibinfo {author} {\bibfnamefont
  {D.~G.}\ \bibnamefont {Cory}}, \bibinfo {author} {\bibfnamefont {M.~G.}\
  \bibnamefont {Huber}}, \bibinfo {author} {\bibfnamefont {C.~A.}\ \bibnamefont
  {Marjerrison}}, \bibinfo {author} {\bibfnamefont {M.}~\bibnamefont {Pula}},
  \bibinfo {author} {\bibfnamefont {D.}~\bibnamefont {Sarenac}}, \emph
  {et~al.},\ }\bibfield  {title} {\bibinfo {title} {Characterization of a
  disordered above room temperature skyrmion material co8zn8mn4},\ }\href@noop
  {} {\bibfield  {journal} {\bibinfo  {journal} {Materials}\ }\textbf {\bibinfo
  {volume} {14}},\ \bibinfo {pages} {4689} (\bibinfo {year}
  {2021})}\BibitemShut {NoStop}%
\bibitem [{\citenamefont {Manyala}\ \emph {et~al.}(2000)\citenamefont
  {Manyala}, \citenamefont {Sidis}, \citenamefont {DiTusa}, \citenamefont
  {Aeppli}, \citenamefont {Young},\ and\ \citenamefont
  {Fisk}}]{manyala2000addendum}%
  \BibitemOpen
  \bibfield  {author} {\bibinfo {author} {\bibfnamefont {N.}~\bibnamefont
  {Manyala}}, \bibinfo {author} {\bibfnamefont {Y.}~\bibnamefont {Sidis}},
  \bibinfo {author} {\bibfnamefont {J.}~\bibnamefont {DiTusa}}, \bibinfo
  {author} {\bibfnamefont {G.}~\bibnamefont {Aeppli}}, \bibinfo {author}
  {\bibfnamefont {D.}~\bibnamefont {Young}},\ and\ \bibinfo {author}
  {\bibfnamefont {Z.}~\bibnamefont {Fisk}},\ }\bibfield  {title} {\bibinfo
  {title} {addendum: Magnetoresistance from quantum interference effects in
  ferromagnets},\ }\href@noop {} {\bibfield  {journal} {\bibinfo  {journal}
  {Nature}\ }\textbf {\bibinfo {volume} {408}},\ \bibinfo {pages} {616}
  (\bibinfo {year} {2000})}\BibitemShut {NoStop}%
\bibitem [{\citenamefont {Rhodes}\ and\ \citenamefont
  {Wohlfarth}(1963)}]{rhodes1963effective}%
  \BibitemOpen
  \bibfield  {author} {\bibinfo {author} {\bibfnamefont {P.}~\bibnamefont
  {Rhodes}}\ and\ \bibinfo {author} {\bibfnamefont {E.~P.}\ \bibnamefont
  {Wohlfarth}},\ }\bibfield  {title} {\bibinfo {title} {The effective
  curie-weiss constant of ferromagnetic metals and alloys},\ }\href@noop {}
  {\bibfield  {journal} {\bibinfo  {journal} {Proceedings of the Royal Society
  of London. Series A. Mathematical and Physical Sciences}\ }\textbf {\bibinfo
  {volume} {273}},\ \bibinfo {pages} {247} (\bibinfo {year}
  {1963})}\BibitemShut {NoStop}%
\bibitem [{\citenamefont {Stickel}(2010)}]{stickel2010data}%
  \BibitemOpen
  \bibfield  {author} {\bibinfo {author} {\bibfnamefont {J.~J.}\ \bibnamefont
  {Stickel}},\ }\bibfield  {title} {\bibinfo {title} {Data smoothing and
  numerical differentiation by a regularization method},\ }\href@noop {}
  {\bibfield  {journal} {\bibinfo  {journal} {Computers \& chemical
  engineering}\ }\textbf {\bibinfo {volume} {34}},\ \bibinfo {pages} {467}
  (\bibinfo {year} {2010})}\BibitemShut {NoStop}%
\bibitem [{\citenamefont {Bocarsly}\ \emph {et~al.}(2018)\citenamefont
  {Bocarsly}, \citenamefont {Need}, \citenamefont {Seshadri},\ and\
  \citenamefont {Wilson}}]{bocarsly2018magnetoentropic}%
  \BibitemOpen
  \bibfield  {author} {\bibinfo {author} {\bibfnamefont {J.~D.}\ \bibnamefont
  {Bocarsly}}, \bibinfo {author} {\bibfnamefont {R.~F.}\ \bibnamefont {Need}},
  \bibinfo {author} {\bibfnamefont {R.}~\bibnamefont {Seshadri}},\ and\
  \bibinfo {author} {\bibfnamefont {S.~D.}\ \bibnamefont {Wilson}},\ }\bibfield
   {title} {\bibinfo {title} {Magnetoentropic signatures of skyrmionic phase
  behavior in fege},\ }\href@noop {} {\bibfield  {journal} {\bibinfo  {journal}
  {Physical Review B}\ }\textbf {\bibinfo {volume} {97}},\ \bibinfo {pages}
  {100404} (\bibinfo {year} {2018})}\BibitemShut {NoStop}%
\bibitem [{\citenamefont {Ahmed}\ \emph {et~al.}(2022)\citenamefont {Ahmed},
  \citenamefont {Mazumdar}, \citenamefont {Das},\ and\ \citenamefont
  {Das}}]{ahmed2022comparative}%
  \BibitemOpen
  \bibfield  {author} {\bibinfo {author} {\bibfnamefont {A.}~\bibnamefont
  {Ahmed}}, \bibinfo {author} {\bibfnamefont {D.}~\bibnamefont {Mazumdar}},
  \bibinfo {author} {\bibfnamefont {K.}~\bibnamefont {Das}},\ and\ \bibinfo
  {author} {\bibfnamefont {I.}~\bibnamefont {Das}},\ }\bibfield  {title}
  {\bibinfo {title} {A comparative study of the magnetic and magnetocaloric
  effect of polycrystalline gd0. 9y0. 1mno3 and gd0. 7y0. 3mno3 compounds:
  Influence of y-ions on the magnetic state of gdmno3},\ }\href@noop {}
  {\bibfield  {journal} {\bibinfo  {journal} {Journal of Magnetism and Magnetic
  Materials}\ }\textbf {\bibinfo {volume} {551}},\ \bibinfo {pages} {169133}
  (\bibinfo {year} {2022})}\BibitemShut {NoStop}%
\bibitem [{\citenamefont {Jamaluddin}\ \emph {et~al.}(2019)\citenamefont
  {Jamaluddin}, \citenamefont {Manna}, \citenamefont {Giri}, \citenamefont
  {Madduri}, \citenamefont {Parkin},\ and\ \citenamefont
  {Nayak}}]{jamaluddin2019robust}%
  \BibitemOpen
  \bibfield  {author} {\bibinfo {author} {\bibfnamefont {S.}~\bibnamefont
  {Jamaluddin}}, \bibinfo {author} {\bibfnamefont {S.~K.}\ \bibnamefont
  {Manna}}, \bibinfo {author} {\bibfnamefont {B.}~\bibnamefont {Giri}},
  \bibinfo {author} {\bibfnamefont {P.~P.}\ \bibnamefont {Madduri}}, \bibinfo
  {author} {\bibfnamefont {S.~S.}\ \bibnamefont {Parkin}},\ and\ \bibinfo
  {author} {\bibfnamefont {A.~K.}\ \bibnamefont {Nayak}},\ }\bibfield  {title}
  {\bibinfo {title} {Robust antiskyrmion phase in bulk tetragonal mn--pt
  (pd)--sn heusler system probed by magnetic entropy change and
  ac-susceptibility measurements},\ }\href@noop {} {\bibfield  {journal}
  {\bibinfo  {journal} {Advanced Functional Materials}\ }\textbf {\bibinfo
  {volume} {29}},\ \bibinfo {pages} {1901776} (\bibinfo {year}
  {2019})}\BibitemShut {NoStop}%
\bibitem [{\citenamefont {Fisher}(1967)}]{sopt1}%
  \BibitemOpen
  \bibfield  {author} {\bibinfo {author} {\bibfnamefont {M.~E.}\ \bibnamefont
  {Fisher}},\ }\bibfield  {title} {\bibinfo {title} {The theory of equilibrium
  critical phenomena},\ }\href {https://doi.org/10.1088/0034-4885/30/2/306}
  {\bibfield  {journal} {\bibinfo  {journal} {Reports on Progress in Physics}\
  }\textbf {\bibinfo {volume} {30}},\ \bibinfo {pages} {615} (\bibinfo {year}
  {1967})}\BibitemShut {NoStop}%
\bibitem [{\citenamefont {Stanley}(1971{\natexlab{a}})}]{sopt2}%
  \BibitemOpen
  \bibfield  {author} {\bibinfo {author} {\bibfnamefont {H.~E.}\ \bibnamefont
  {Stanley}},\ }\href@noop {} {\emph {\bibinfo {title} {Phase transitions and
  critical phenomena}}},\ Vol.~\bibinfo {volume} {7}\ (\bibinfo  {publisher}
  {Clarendon Press, Oxford},\ \bibinfo {year} {1971})\BibitemShut {NoStop}%
\bibitem [{\citenamefont {Banerjee}(1964)}]{Banerjee}%
  \BibitemOpen
  \bibfield  {author} {\bibinfo {author} {\bibfnamefont {B.}~\bibnamefont
  {Banerjee}},\ }\bibfield  {title} {\bibinfo {title} {On a generalised
  approach to first and second order magnetic transitions},\ }\href
  {https://doi.org/https://doi.org/10.1016/0031-9163(64)91158-8} {\bibfield
  {journal} {\bibinfo  {journal} {Physics Letters}\ }\textbf {\bibinfo {volume}
  {12}},\ \bibinfo {pages} {16} (\bibinfo {year} {1964})}\BibitemShut {NoStop}%
\bibitem [{\citenamefont {Arrott}\ and\ \citenamefont
  {Noakes}(1967)}]{arrotnoakes}%
  \BibitemOpen
  \bibfield  {author} {\bibinfo {author} {\bibfnamefont {A.}~\bibnamefont
  {Arrott}}\ and\ \bibinfo {author} {\bibfnamefont {J.~E.}\ \bibnamefont
  {Noakes}},\ }\bibfield  {title} {\bibinfo {title} {Approximate equation of
  state for nickel near its critical temperature},\ }\href
  {https://doi.org/10.1103/PhysRevLett.19.786} {\bibfield  {journal} {\bibinfo
  {journal} {Phys. Rev. Lett.}\ }\textbf {\bibinfo {volume} {19}},\ \bibinfo
  {pages} {786} (\bibinfo {year} {1967})}\BibitemShut {NoStop}%
\bibitem [{\citenamefont {Kaul}(1985)}]{theory}%
  \BibitemOpen
  \bibfield  {author} {\bibinfo {author} {\bibfnamefont {S.}~\bibnamefont
  {Kaul}},\ }\bibfield  {title} {\bibinfo {title} {Static critical phenomena in
  ferromagnets with quenched disorder},\ }\href@noop {} {\bibfield  {journal}
  {\bibinfo  {journal} {Journal of magnetism and magnetic materials}\ }\textbf
  {\bibinfo {volume} {53}},\ \bibinfo {pages} {5} (\bibinfo {year}
  {1985})}\BibitemShut {NoStop}%
\bibitem [{\citenamefont {Huang}(2009)}]{theorytcp}%
  \BibitemOpen
  \bibfield  {author} {\bibinfo {author} {\bibfnamefont {K.}~\bibnamefont
  {Huang}},\ }\href@noop {} {\emph {\bibinfo {title} {Introduction to
  statistical physics}}}\ (\bibinfo  {publisher} {Chapman and Hall/CRC},\
  \bibinfo {year} {2009})\BibitemShut {NoStop}%
\bibitem [{\citenamefont {Fan}\ \emph {et~al.}(2010)\citenamefont {Fan},
  \citenamefont {Ling}, \citenamefont {Hong}, \citenamefont {Zhang},
  \citenamefont {Pi},\ and\ \citenamefont {Zhang}}]{ns}%
  \BibitemOpen
  \bibfield  {author} {\bibinfo {author} {\bibfnamefont {J.}~\bibnamefont
  {Fan}}, \bibinfo {author} {\bibfnamefont {L.}~\bibnamefont {Ling}}, \bibinfo
  {author} {\bibfnamefont {B.}~\bibnamefont {Hong}}, \bibinfo {author}
  {\bibfnamefont {L.}~\bibnamefont {Zhang}}, \bibinfo {author} {\bibfnamefont
  {L.}~\bibnamefont {Pi}},\ and\ \bibinfo {author} {\bibfnamefont
  {Y.}~\bibnamefont {Zhang}},\ }\bibfield  {title} {\bibinfo {title} {Critical
  properties of the perovskite manganite
  ${\text{la}}_{0.1}{\text{nd}}_{0.6}{\text{sr}}_{0.3}{\text{mno}}_{3}$},\
  }\href {https://doi.org/10.1103/PhysRevB.81.144426} {\bibfield  {journal}
  {\bibinfo  {journal} {Phys. Rev. B}\ }\textbf {\bibinfo {volume} {81}},\
  \bibinfo {pages} {144426} (\bibinfo {year} {2010})}\BibitemShut {NoStop}%
\bibitem [{\citenamefont {Kouvel}\ and\ \citenamefont {Fisher}(1964)}]{KF}%
  \BibitemOpen
  \bibfield  {author} {\bibinfo {author} {\bibfnamefont {J.~S.}\ \bibnamefont
  {Kouvel}}\ and\ \bibinfo {author} {\bibfnamefont {M.~E.}\ \bibnamefont
  {Fisher}},\ }\bibfield  {title} {\bibinfo {title} {Detailed magnetic behavior
  of nickel near its curie point},\ }\href@noop {} {\bibfield  {journal}
  {\bibinfo  {journal} {Physical Review}\ }\textbf {\bibinfo {volume} {136}},\
  \bibinfo {pages} {A1626} (\bibinfo {year} {1964})}\BibitemShut {NoStop}%
\bibitem [{\citenamefont {Kadanoff}(1966)}]{delta}%
  \BibitemOpen
  \bibfield  {author} {\bibinfo {author} {\bibfnamefont {L.~P.}\ \bibnamefont
  {Kadanoff}},\ }\bibfield  {title} {\bibinfo {title} {Scaling laws for ising
  models near t c},\ }\href@noop {} {\bibfield  {journal} {\bibinfo  {journal}
  {Physics Physique Fizika}\ }\textbf {\bibinfo {volume} {2}},\ \bibinfo
  {pages} {263} (\bibinfo {year} {1966})}\BibitemShut {NoStop}%
\bibitem [{\citenamefont {Stanley}(1971{\natexlab{b}})}]{scaling}%
  \BibitemOpen
  \bibfield  {author} {\bibinfo {author} {\bibfnamefont {H.~E.}\ \bibnamefont
  {Stanley}},\ }\href@noop {} {\emph {\bibinfo {title} {Phase transitions and
  critical phenomena}}},\ Vol.~\bibinfo {volume} {7}\ (\bibinfo  {publisher}
  {Clarendon Press, Oxford},\ \bibinfo {year} {1971})\BibitemShut {NoStop}%
\bibitem [{\citenamefont {Hurd}(2012)}]{hall1}%
  \BibitemOpen
  \bibfield  {author} {\bibinfo {author} {\bibfnamefont {C.}~\bibnamefont
  {Hurd}},\ }\href@noop {} {\emph {\bibinfo {title} {The Hall effect in metals
  and alloys}}}\ (\bibinfo  {publisher} {Springer Science \& Business Media},\
  \bibinfo {year} {2012})\BibitemShut {NoStop}%
\bibitem [{\citenamefont {Taguchi}\ \emph {et~al.}(2001)\citenamefont
  {Taguchi}, \citenamefont {Oohara}, \citenamefont {Yoshizawa}, \citenamefont
  {Nagaosa},\ and\ \citenamefont {Tokura}}]{taguchi2001spin}%
  \BibitemOpen
  \bibfield  {author} {\bibinfo {author} {\bibfnamefont {Y.}~\bibnamefont
  {Taguchi}}, \bibinfo {author} {\bibfnamefont {Y.}~\bibnamefont {Oohara}},
  \bibinfo {author} {\bibfnamefont {H.}~\bibnamefont {Yoshizawa}}, \bibinfo
  {author} {\bibfnamefont {N.}~\bibnamefont {Nagaosa}},\ and\ \bibinfo {author}
  {\bibfnamefont {Y.}~\bibnamefont {Tokura}},\ }\bibfield  {title} {\bibinfo
  {title} {Spin chirality, berry phase, and anomalous hall effect in a
  frustrated ferromagnet},\ }\href@noop {} {\bibfield  {journal} {\bibinfo
  {journal} {Science}\ }\textbf {\bibinfo {volume} {291}},\ \bibinfo {pages}
  {2573} (\bibinfo {year} {2001})}\BibitemShut {NoStop}%
\bibitem [{\citenamefont {Lee}\ \emph {et~al.}(2004)\citenamefont {Lee},
  \citenamefont {Watauchi}, \citenamefont {Miller}, \citenamefont {Cava},\ and\
  \citenamefont {Ong}}]{lee2004dissipationless}%
  \BibitemOpen
  \bibfield  {author} {\bibinfo {author} {\bibfnamefont {W.-L.}\ \bibnamefont
  {Lee}}, \bibinfo {author} {\bibfnamefont {S.}~\bibnamefont {Watauchi}},
  \bibinfo {author} {\bibfnamefont {V.}~\bibnamefont {Miller}}, \bibinfo
  {author} {\bibfnamefont {R.}~\bibnamefont {Cava}},\ and\ \bibinfo {author}
  {\bibfnamefont {N.}~\bibnamefont {Ong}},\ }\bibfield  {title} {\bibinfo
  {title} {Dissipationless anomalous hall current in the ferromagnetic spinel
  cucr2se4-x br x},\ }\href@noop {} {\bibfield  {journal} {\bibinfo  {journal}
  {Science}\ }\textbf {\bibinfo {volume} {303}},\ \bibinfo {pages} {1647}
  (\bibinfo {year} {2004})}\BibitemShut {NoStop}%
\bibitem [{\citenamefont {Paschen}\ \emph {et~al.}(2004)\citenamefont
  {Paschen}, \citenamefont {L{\"u}hmann}, \citenamefont {Wirth}, \citenamefont
  {Gegenwart}, \citenamefont {Trovarelli}, \citenamefont {Geibel},
  \citenamefont {Steglich}, \citenamefont {Coleman},\ and\ \citenamefont
  {Si}}]{paschen2004hall}%
  \BibitemOpen
  \bibfield  {author} {\bibinfo {author} {\bibfnamefont {S.}~\bibnamefont
  {Paschen}}, \bibinfo {author} {\bibfnamefont {T.}~\bibnamefont
  {L{\"u}hmann}}, \bibinfo {author} {\bibfnamefont {S.}~\bibnamefont {Wirth}},
  \bibinfo {author} {\bibfnamefont {P.}~\bibnamefont {Gegenwart}}, \bibinfo
  {author} {\bibfnamefont {O.}~\bibnamefont {Trovarelli}}, \bibinfo {author}
  {\bibfnamefont {C.}~\bibnamefont {Geibel}}, \bibinfo {author} {\bibfnamefont
  {F.}~\bibnamefont {Steglich}}, \bibinfo {author} {\bibfnamefont
  {P.}~\bibnamefont {Coleman}},\ and\ \bibinfo {author} {\bibfnamefont
  {Q.}~\bibnamefont {Si}},\ }\bibfield  {title} {\bibinfo {title} {Hall-effect
  evolution across a heavy-fermion quantum critical point},\ }\href@noop {}
  {\bibfield  {journal} {\bibinfo  {journal} {Nature}\ }\textbf {\bibinfo
  {volume} {432}},\ \bibinfo {pages} {881} (\bibinfo {year}
  {2004})}\BibitemShut {NoStop}%
\bibitem [{\citenamefont {Miyasato}\ \emph {et~al.}(2007)\citenamefont
  {Miyasato}, \citenamefont {Abe}, \citenamefont {Fujii}, \citenamefont
  {Asamitsu}, \citenamefont {Onoda}, \citenamefont {Onose}, \citenamefont
  {Nagaosa},\ and\ \citenamefont {Tokura}}]{miyasato2007crossover}%
  \BibitemOpen
  \bibfield  {author} {\bibinfo {author} {\bibfnamefont {T.}~\bibnamefont
  {Miyasato}}, \bibinfo {author} {\bibfnamefont {N.}~\bibnamefont {Abe}},
  \bibinfo {author} {\bibfnamefont {T.}~\bibnamefont {Fujii}}, \bibinfo
  {author} {\bibfnamefont {A.}~\bibnamefont {Asamitsu}}, \bibinfo {author}
  {\bibfnamefont {S.}~\bibnamefont {Onoda}}, \bibinfo {author} {\bibfnamefont
  {Y.}~\bibnamefont {Onose}}, \bibinfo {author} {\bibfnamefont
  {N.}~\bibnamefont {Nagaosa}},\ and\ \bibinfo {author} {\bibfnamefont
  {Y.}~\bibnamefont {Tokura}},\ }\bibfield  {title} {\bibinfo {title}
  {Crossover behavior of the anomalous hall effect and anomalous nernst effect
  in itinerant ferromagnets},\ }\href@noop {} {\bibfield  {journal} {\bibinfo
  {journal} {Physical review letters}\ }\textbf {\bibinfo {volume} {99}},\
  \bibinfo {pages} {086602} (\bibinfo {year} {2007})}\BibitemShut {NoStop}%
\bibitem [{\citenamefont {Fang}\ \emph {et~al.}(2003)\citenamefont {Fang},
  \citenamefont {Nagaosa}, \citenamefont {Takahashi}, \citenamefont {Asamitsu},
  \citenamefont {Mathieu}, \citenamefont {Ogasawara}, \citenamefont {Yamada},
  \citenamefont {Kawasaki}, \citenamefont {Tokura},\ and\ \citenamefont
  {Terakura}}]{fang2003anomalous}%
  \BibitemOpen
  \bibfield  {author} {\bibinfo {author} {\bibfnamefont {Z.}~\bibnamefont
  {Fang}}, \bibinfo {author} {\bibfnamefont {N.}~\bibnamefont {Nagaosa}},
  \bibinfo {author} {\bibfnamefont {K.~S.}\ \bibnamefont {Takahashi}}, \bibinfo
  {author} {\bibfnamefont {A.}~\bibnamefont {Asamitsu}}, \bibinfo {author}
  {\bibfnamefont {R.}~\bibnamefont {Mathieu}}, \bibinfo {author} {\bibfnamefont
  {T.}~\bibnamefont {Ogasawara}}, \bibinfo {author} {\bibfnamefont
  {H.}~\bibnamefont {Yamada}}, \bibinfo {author} {\bibfnamefont
  {M.}~\bibnamefont {Kawasaki}}, \bibinfo {author} {\bibfnamefont
  {Y.}~\bibnamefont {Tokura}},\ and\ \bibinfo {author} {\bibfnamefont
  {K.}~\bibnamefont {Terakura}},\ }\bibfield  {title} {\bibinfo {title} {The
  anomalous hall effect and magnetic monopoles in momentum space},\ }\href@noop
  {} {\bibfield  {journal} {\bibinfo  {journal} {Science}\ }\textbf {\bibinfo
  {volume} {302}},\ \bibinfo {pages} {92} (\bibinfo {year} {2003})}\BibitemShut
  {NoStop}%
\bibitem [{\citenamefont {Tian}\ \emph {et~al.}(2009)\citenamefont {Tian},
  \citenamefont {Ye},\ and\ \citenamefont {Jin}}]{tian2009proper}%
  \BibitemOpen
  \bibfield  {author} {\bibinfo {author} {\bibfnamefont {Y.}~\bibnamefont
  {Tian}}, \bibinfo {author} {\bibfnamefont {L.}~\bibnamefont {Ye}},\ and\
  \bibinfo {author} {\bibfnamefont {X.}~\bibnamefont {Jin}},\ }\bibfield
  {title} {\bibinfo {title} {Proper scaling of the anomalous hall effect},\
  }\href@noop {} {\bibfield  {journal} {\bibinfo  {journal} {Physical review
  letters}\ }\textbf {\bibinfo {volume} {103}},\ \bibinfo {pages} {087206}
  (\bibinfo {year} {2009})}\BibitemShut {NoStop}%
\bibitem [{\citenamefont {Karplus}\ and\ \citenamefont
  {Luttinger}(1954)}]{karplus1954hall}%
  \BibitemOpen
  \bibfield  {author} {\bibinfo {author} {\bibfnamefont {R.}~\bibnamefont
  {Karplus}}\ and\ \bibinfo {author} {\bibfnamefont {J.}~\bibnamefont
  {Luttinger}},\ }\bibfield  {title} {\bibinfo {title} {Hall effect in
  ferromagnetics},\ }\href@noop {} {\bibfield  {journal} {\bibinfo  {journal}
  {Physical Review}\ }\textbf {\bibinfo {volume} {95}},\ \bibinfo {pages}
  {1154} (\bibinfo {year} {1954})}\BibitemShut {NoStop}%
\bibitem [{\citenamefont {Husmann}\ and\ \citenamefont
  {Singh}(2006)}]{husmann2006temperature}%
  \BibitemOpen
  \bibfield  {author} {\bibinfo {author} {\bibfnamefont {A.}~\bibnamefont
  {Husmann}}\ and\ \bibinfo {author} {\bibfnamefont {L.}~\bibnamefont
  {Singh}},\ }\bibfield  {title} {\bibinfo {title} {Temperature dependence of
  the anomalous hall conductivity in the heusler alloy co 2 cr al},\
  }\href@noop {} {\bibfield  {journal} {\bibinfo  {journal} {Physical Review
  B}\ }\textbf {\bibinfo {volume} {73}},\ \bibinfo {pages} {172417} (\bibinfo
  {year} {2006})}\BibitemShut {NoStop}%
\bibitem [{\citenamefont {Lavine}(1961)}]{lavine1961extraordinary}%
  \BibitemOpen
  \bibfield  {author} {\bibinfo {author} {\bibfnamefont {J.~M.}\ \bibnamefont
  {Lavine}},\ }\bibfield  {title} {\bibinfo {title} {Extraordinary hall-effect
  measurements on ni, some ni alloys, and ferrites},\ }\href@noop {} {\bibfield
   {journal} {\bibinfo  {journal} {Physical Review}\ }\textbf {\bibinfo
  {volume} {123}},\ \bibinfo {pages} {1273} (\bibinfo {year}
  {1961})}\BibitemShut {NoStop}%
\bibitem [{\citenamefont {Smit}(1955)}]{smit1955spontaneous}%
  \BibitemOpen
  \bibfield  {author} {\bibinfo {author} {\bibfnamefont {J.}~\bibnamefont
  {Smit}},\ }\bibfield  {title} {\bibinfo {title} {The spontaneous hall effect
  in ferromagnetics i},\ }\href@noop {} {\bibfield  {journal} {\bibinfo
  {journal} {Physica}\ }\textbf {\bibinfo {volume} {21}},\ \bibinfo {pages}
  {877} (\bibinfo {year} {1955})}\BibitemShut {NoStop}%
\bibitem [{\citenamefont {Berger}(1970)}]{berger1970side}%
  \BibitemOpen
  \bibfield  {author} {\bibinfo {author} {\bibfnamefont {L.}~\bibnamefont
  {Berger}},\ }\bibfield  {title} {\bibinfo {title} {Side-jump mechanism for
  the hall effect of ferromagnets},\ }\href@noop {} {\bibfield  {journal}
  {\bibinfo  {journal} {Physical Review B}\ }\textbf {\bibinfo {volume} {2}},\
  \bibinfo {pages} {4559} (\bibinfo {year} {1970})}\BibitemShut {NoStop}%
\bibitem [{\citenamefont {Onoda}\ \emph {et~al.}(2006)\citenamefont {Onoda},
  \citenamefont {Sugimoto},\ and\ \citenamefont
  {Nagaosa}}]{onoda2006intrinsic}%
  \BibitemOpen
  \bibfield  {author} {\bibinfo {author} {\bibfnamefont {S.}~\bibnamefont
  {Onoda}}, \bibinfo {author} {\bibfnamefont {N.}~\bibnamefont {Sugimoto}},\
  and\ \bibinfo {author} {\bibfnamefont {N.}~\bibnamefont {Nagaosa}},\
  }\bibfield  {title} {\bibinfo {title} {Intrinsic versus extrinsic anomalous
  hall effect in ferromagnets},\ }\href@noop {} {\bibfield  {journal} {\bibinfo
   {journal} {Physical review letters}\ }\textbf {\bibinfo {volume} {97}},\
  \bibinfo {pages} {126602} (\bibinfo {year} {2006})}\BibitemShut {NoStop}%
\bibitem [{\citenamefont {Zeng}\ \emph {et~al.}(2006)\citenamefont {Zeng},
  \citenamefont {Yao}, \citenamefont {Niu},\ and\ \citenamefont
  {Weitering}}]{zeng2006linear}%
  \BibitemOpen
  \bibfield  {author} {\bibinfo {author} {\bibfnamefont {C.}~\bibnamefont
  {Zeng}}, \bibinfo {author} {\bibfnamefont {Y.}~\bibnamefont {Yao}}, \bibinfo
  {author} {\bibfnamefont {Q.}~\bibnamefont {Niu}},\ and\ \bibinfo {author}
  {\bibfnamefont {H.~H.}\ \bibnamefont {Weitering}},\ }\bibfield  {title}
  {\bibinfo {title} {Linear magnetization dependence of the intrinsic anomalous
  hall effect},\ }\href@noop {} {\bibfield  {journal} {\bibinfo  {journal}
  {Physical review letters}\ }\textbf {\bibinfo {volume} {96}},\ \bibinfo
  {pages} {037204} (\bibinfo {year} {2006})}\BibitemShut {NoStop}%
\bibitem [{\citenamefont {Jungwirth}\ \emph {et~al.}(2002)\citenamefont
  {Jungwirth}, \citenamefont {Niu},\ and\ \citenamefont
  {MacDonald}}]{jungwirth2002anomalous}%
  \BibitemOpen
  \bibfield  {author} {\bibinfo {author} {\bibfnamefont {T.}~\bibnamefont
  {Jungwirth}}, \bibinfo {author} {\bibfnamefont {Q.}~\bibnamefont {Niu}},\
  and\ \bibinfo {author} {\bibfnamefont {A.}~\bibnamefont {MacDonald}},\
  }\bibfield  {title} {\bibinfo {title} {Anomalous hall effect in ferromagnetic
  semiconductors},\ }\href@noop {} {\bibfield  {journal} {\bibinfo  {journal}
  {Physical review letters}\ }\textbf {\bibinfo {volume} {88}},\ \bibinfo
  {pages} {207208} (\bibinfo {year} {2002})}\BibitemShut {NoStop}%
\bibitem [{\citenamefont {Jiang}\ \emph {et~al.}(2010)\citenamefont {Jiang},
  \citenamefont {Zhou},\ and\ \citenamefont {Williams}}]{jiang2010scaling}%
  \BibitemOpen
  \bibfield  {author} {\bibinfo {author} {\bibfnamefont {W.}~\bibnamefont
  {Jiang}}, \bibinfo {author} {\bibfnamefont {X.}~\bibnamefont {Zhou}},\ and\
  \bibinfo {author} {\bibfnamefont {G.}~\bibnamefont {Williams}},\ }\bibfield
  {title} {\bibinfo {title} {Scaling the anomalous hall effect: A connection
  between transport and magnetism},\ }\href@noop {} {\bibfield  {journal}
  {\bibinfo  {journal} {Physical Review B}\ }\textbf {\bibinfo {volume} {82}},\
  \bibinfo {pages} {144424} (\bibinfo {year} {2010})}\BibitemShut {NoStop}%
\bibitem [{\citenamefont {Jiang}\ \emph {et~al.}(2009)\citenamefont {Jiang},
  \citenamefont {Wirthmann}, \citenamefont {Gui}, \citenamefont {Zhou},
  \citenamefont {Reinwald}, \citenamefont {Wegscheider}, \citenamefont {Hu},\
  and\ \citenamefont {Williams}}]{jiang2009critical}%
  \BibitemOpen
  \bibfield  {author} {\bibinfo {author} {\bibfnamefont {W.}~\bibnamefont
  {Jiang}}, \bibinfo {author} {\bibfnamefont {A.}~\bibnamefont {Wirthmann}},
  \bibinfo {author} {\bibfnamefont {Y.}~\bibnamefont {Gui}}, \bibinfo {author}
  {\bibfnamefont {X.}~\bibnamefont {Zhou}}, \bibinfo {author} {\bibfnamefont
  {M.}~\bibnamefont {Reinwald}}, \bibinfo {author} {\bibfnamefont
  {W.}~\bibnamefont {Wegscheider}}, \bibinfo {author} {\bibfnamefont {C.-M.}\
  \bibnamefont {Hu}},\ and\ \bibinfo {author} {\bibfnamefont {G.}~\bibnamefont
  {Williams}},\ }\bibfield  {title} {\bibinfo {title} {Critical behavior from
  the anomalous hall effect in (gamn) as},\ }\href@noop {} {\bibfield
  {journal} {\bibinfo  {journal} {Physical Review B}\ }\textbf {\bibinfo
  {volume} {80}},\ \bibinfo {pages} {214409} (\bibinfo {year}
  {2009})}\BibitemShut {NoStop}%
\bibitem [{\citenamefont {Fisher}\ \emph {et~al.}(1972)\citenamefont {Fisher},
  \citenamefont {Ma},\ and\ \citenamefont {Nickel}}]{fisher1972critical}%
  \BibitemOpen
  \bibfield  {author} {\bibinfo {author} {\bibfnamefont {M.~E.}\ \bibnamefont
  {Fisher}}, \bibinfo {author} {\bibfnamefont {S.-k.}\ \bibnamefont {Ma}},\
  and\ \bibinfo {author} {\bibfnamefont {B.}~\bibnamefont {Nickel}},\
  }\bibfield  {title} {\bibinfo {title} {Critical exponents for long-range
  interactions},\ }\href@noop {} {\bibfield  {journal} {\bibinfo  {journal}
  {Physical Review Letters}\ }\textbf {\bibinfo {volume} {29}},\ \bibinfo
  {pages} {917} (\bibinfo {year} {1972})}\BibitemShut {NoStop}%
\bibitem [{\citenamefont {Wilhelm}\ \emph {et~al.}(2011)\citenamefont
  {Wilhelm}, \citenamefont {Baenitz}, \citenamefont {Schmidt}, \citenamefont
  {R{\"o}{\ss}ler}, \citenamefont {Leonov},\ and\ \citenamefont
  {Bogdanov}}]{wilhelm2011precursor}%
  \BibitemOpen
  \bibfield  {author} {\bibinfo {author} {\bibfnamefont {H.}~\bibnamefont
  {Wilhelm}}, \bibinfo {author} {\bibfnamefont {M.}~\bibnamefont {Baenitz}},
  \bibinfo {author} {\bibfnamefont {M.}~\bibnamefont {Schmidt}}, \bibinfo
  {author} {\bibfnamefont {U.}~\bibnamefont {R{\"o}{\ss}ler}}, \bibinfo
  {author} {\bibfnamefont {A.}~\bibnamefont {Leonov}},\ and\ \bibinfo {author}
  {\bibfnamefont {A.}~\bibnamefont {Bogdanov}},\ }\bibfield  {title} {\bibinfo
  {title} {Precursor phenomena at the magnetic ordering of the cubic helimagnet
  fege},\ }\href@noop {} {\bibfield  {journal} {\bibinfo  {journal} {Physical
  review letters}\ }\textbf {\bibinfo {volume} {107}},\ \bibinfo {pages}
  {127203} (\bibinfo {year} {2011})}\BibitemShut {NoStop}%
\bibitem [{\citenamefont {Wilhelm}\ \emph {et~al.}(2012)\citenamefont
  {Wilhelm}, \citenamefont {Baenitz}, \citenamefont {Schmidt}, \citenamefont
  {Naylor}, \citenamefont {Lortz}, \citenamefont {R{\"o}{\ss}ler},
  \citenamefont {Leonov},\ and\ \citenamefont
  {Bogdanov}}]{wilhelm2012confinement}%
  \BibitemOpen
  \bibfield  {author} {\bibinfo {author} {\bibfnamefont {H.}~\bibnamefont
  {Wilhelm}}, \bibinfo {author} {\bibfnamefont {M.}~\bibnamefont {Baenitz}},
  \bibinfo {author} {\bibfnamefont {M.}~\bibnamefont {Schmidt}}, \bibinfo
  {author} {\bibfnamefont {C.}~\bibnamefont {Naylor}}, \bibinfo {author}
  {\bibfnamefont {R.}~\bibnamefont {Lortz}}, \bibinfo {author} {\bibfnamefont
  {U.}~\bibnamefont {R{\"o}{\ss}ler}}, \bibinfo {author} {\bibfnamefont
  {A.}~\bibnamefont {Leonov}},\ and\ \bibinfo {author} {\bibfnamefont
  {A.}~\bibnamefont {Bogdanov}},\ }\bibfield  {title} {\bibinfo {title}
  {Confinement of chiral magnetic modulations in the precursor region of
  fege},\ }\href@noop {} {\bibfield  {journal} {\bibinfo  {journal} {Journal of
  Physics: Condensed Matter}\ }\textbf {\bibinfo {volume} {24}},\ \bibinfo
  {pages} {294204} (\bibinfo {year} {2012})}\BibitemShut {NoStop}%
\bibitem [{\citenamefont {Shen}\ \emph {et~al.}(2002)\citenamefont {Shen},
  \citenamefont {Schwarz}, \citenamefont {Coulter},\ and\ \citenamefont
  {Thompson}}]{shen2002magnetocaloric}%
  \BibitemOpen
  \bibfield  {author} {\bibinfo {author} {\bibfnamefont {T.}~\bibnamefont
  {Shen}}, \bibinfo {author} {\bibfnamefont {R.}~\bibnamefont {Schwarz}},
  \bibinfo {author} {\bibfnamefont {J.}~\bibnamefont {Coulter}},\ and\ \bibinfo
  {author} {\bibfnamefont {J.}~\bibnamefont {Thompson}},\ }\bibfield  {title}
  {\bibinfo {title} {Magnetocaloric effect in bulk amorphous pd 40 ni 22.5 fe
  17.5 p 20 alloy},\ }\href@noop {} {\bibfield  {journal} {\bibinfo  {journal}
  {Journal of applied physics}\ }\textbf {\bibinfo {volume} {91}},\ \bibinfo
  {pages} {5240} (\bibinfo {year} {2002})}\BibitemShut {NoStop}%
\end{thebibliography}%
\end{document}